\documentclass[lettersize,journal]{IEEEtran}
\usepackage{array}
\usepackage{textcomp}
\usepackage{stfloats}
\usepackage{url}
\usepackage{verbatim}
\usepackage{graphicx}
\usepackage{cite}
\usepackage{color}
\usepackage{xcolor}
\usepackage{graphicx}
\usepackage{subfigure}
\usepackage{multirow}
\usepackage{epstopdf}
\usepackage[ruled,vlined]{algorithm2e}
 \usepackage{makecell}
\usepackage{setspace}
\usepackage{algpseudocode}
\usepackage{amsthm,amsmath,amsfonts,amssymb}
\usepackage{balance}
\usepackage{footmisc}
\usepackage{tabularx}
\usepackage{booktabs}
 \newtheorem{theorem}{Theorem}[]
  \newtheorem{lemma}[theorem]{Lemma}

  \newtheorem{definition}[theorem]{Definition}

\newcommand{\rev}[1]{{#1}}

\begin{document}
\title{DIMS: Distributed Index for Similarity Search in Metric Spaces}

\vspace{-0.5cm}
\author{Yifan~Zhu,
        Chengyang~Luo,
        Tang~Qian,
        Lu~Chen,
        Yunjun~Gao, ~\IEEEmembership{Senior Member,~IEEE,}
        Baihua~Zheng

\thanks{
Y. Zhu, C. Luo, T. Qian, L. Chen and Y. Gao (Corresponding Author) are with the College of Computer Science, Zhejiang University, Hangzhou 310027, China, E-mail:\{xtf\_z, luocy1017, qt.tang.qian, luchen, gaoyj\}@zju.edu.cn.}
\thanks{
B. Zheng is with the School of Computing and Information Systems, Singapore Management University, Singapore,  E-mail: bhzheng@smu.edu.sg.
}}

\maketitle


\begin{abstract}
Similarity search finds objects that are similar to a given query object based on a similarity metric. As the amount and variety of data continue to grow, similarity search in metric spaces has gained significant attention. Metric spaces can accommodate any type of data and support flexible distance metrics, making similarity search in metric spaces beneficial for many real-world applications, such as multimedia retrieval, personalized recommendation, trajectory analytics, \rev{data mining, decision planning, and distributed servers}. However, existing studies mostly focus on indexing metric spaces on a single machine, which faces efficiency and scalability limitations with increasing data volume and query amount. \rev{Recent advancements in similarity search turn towards distributed methods, while they face challenges including inefficient local data management, unbalanced workload, and low concurrent search efficiency.}
To this end, we propose \textbf{DIMS}, an efficient \textbf{D}istributed \textbf{I}ndex for similarity search in \textbf{M}etric \textbf{S}paces. First, we design a novel three-stage heterogeneous partition to achieve workload balance. Then, we present an effective three-stage indexing structure to efficiently manage objects. We also develop concurrent search methods with filtering and validation techniques that support efficient distributed similarity search. Additionally, we devise a cost-based optimization model to balance communication and computation cost. Extensive experiments demonstrate that DIMS significantly outperforms existing distributed similarity search approaches.
\end{abstract}



\section{Introduction}
\label{sec:intro}

%
\IEEEPARstart{T}he proliferation and rapid development of IoT have led to an unprecedented amount of data being generated every day. For example, more than 500 million tweets are posted daily, each containing a variety of data types, including locations, text, and images~\cite{antonakaki2021survey}. To manage this massive volume of various data, there is an urgent need for a general model to store and manage such data. Metric space provides a general solution to accommodate data of different types and volumes, while also supporting flexible distance metrics. As a result, similarity search in metric space has gained significant attention in recent years and offers substantial benefits to a wide range of applications, including multimedia retrieval, personalized recommendation, trajectory analytics, \rev{data mining, decision planning, and distributed servers}~\cite{ fcsc/GouDWK24,vldb/ZhangLZQZ19,mm/YuCLHLW22,sigir/CaoCLW22,kdd/Fang0ZHCGJ22,pvldb/ChatzakisFKPP23,jiis/MaCHH18,sigmod/ZengT022}.


Existing studies on metric space indexing include
compact partitioning methods~\cite{prl/ChavezN05,is/ChavezLRR16,tse/KalantariM83,ipl/Uhlmann91,vldb/CiacciaPZ97}, pivot-based methods~\cite{prl/MicoOV94,tods/BozkayaO99,tkde/ChenGLJC17}, and hybrid methods~\cite{vldb/Brin95,is/NovakBZ11}. However, their focus has predominantly been on single-machine solutions, which often encounter performance bottlenecks due to limited in-memory storage capacity.
To address this challenge, metric indexes often store data on disks, resulting in high I/O costs during similarity search operations. Additionally, the proliferation of online services has led to an influx of simultaneous query requests in data management systems.
For instance, Google's database stores 100PB of data~\cite{urlgoogle} and handles over 2.4 million queries per minute~\cite{urlwwws}.
Single-machine methods struggle to index such vast data volumes or meet the demanding query requirements for such high throughput.
Therefore, there is an urgent need for large-scale similarity search solutions in distributed environments that can efficiently handle large volumes of data and query requests~\cite{dase/UenoSMFM17}.



To address this limitation, various distributed approaches have been proposed. These methods fall into two categories: 
distributed indexes tailored for specified metric spaces and adaptations of existing single machine metric indexes for distributed environments. The former typically leverages a global index with local indexes to support efficient distributed similarity search~\cite{shang2018dita,xie2017distributed,tkde/YagoubiAMP20,icde/ZhengWZZ0J21, www/ZhangGZCWG20}. 
\rev{
However, these methods are designed for specified data types, rendering them inefficient for indexing data of diverse types.}
For example, trie-like distributed indexes used in trajectory analytics~\cite{shang2018dita,icde/ZhengWZZ0J21} often outperform distributed R-tree~\cite{jidm/OliveiraFRCC15}, which is commonly used for high-dimension vector data. Nevertheless, both approaches exhibit inefficiency when dealing with the general metric space that can accommodate various data types.
Thus, existing distributed global and local indexes prove inadequate for effectively modeling the general metric space.

\begin{figure}
  \includegraphics[width=0.99\linewidth]{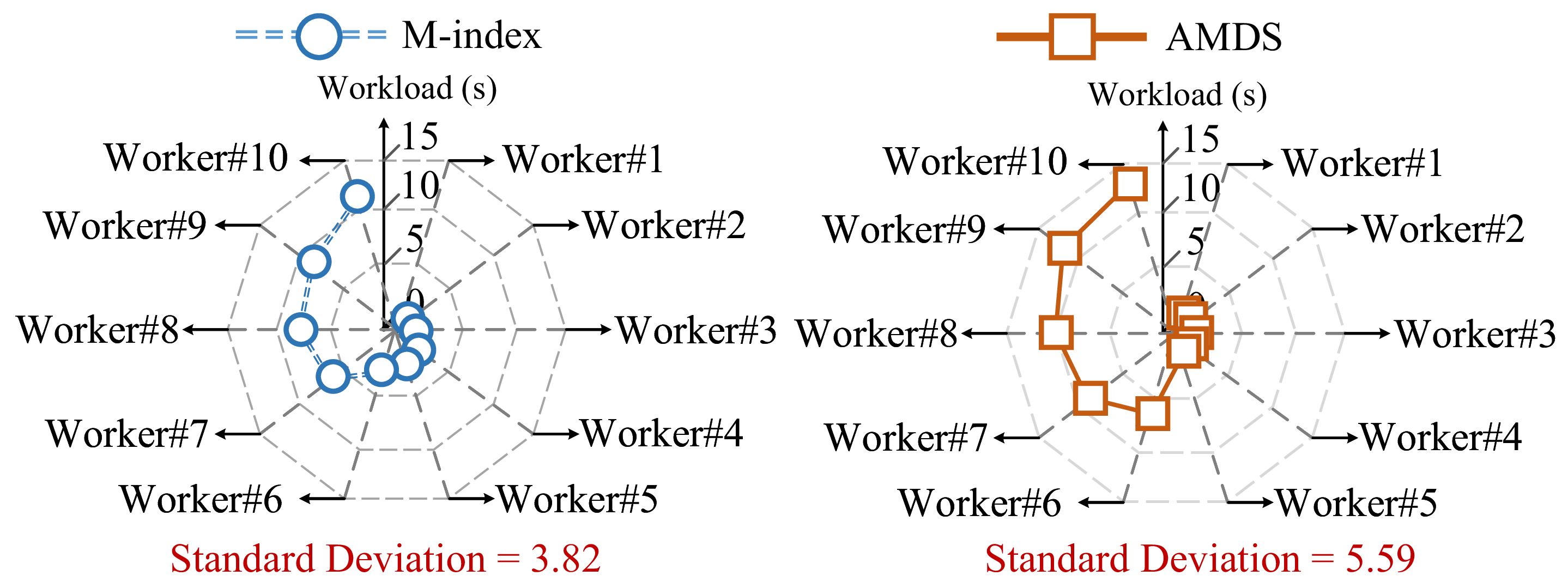}
  \vspace{-3mm}
  \caption{{The imbalanced real workloads}}
  \vspace{-5mm}
  \label{fig:motiv}
\end{figure}

The latter implements existing single machine metric indexes in distributed environments by partitioning objects with pivots or iDistance~\cite{tods/JagadishOTYZ05}, and managing objects in worker nodes with metric indexes.
However, these implementations fail to leverage the full potential of the global index to capture data characteristics, which limits their global pruning power. Furthermore, these approaches typically rely on a homogeneous partition strategy for objects distribution, leading to workload imbalance problems and underutilization of computation resources. For instance, statistical results reported in Fig.~\ref{fig:motiv} illustrate the workload distribution of existing methods M-index~\cite{waim/ZhuSKNY12} and AMDS~\cite{ dase/YangDZCZG19} among ten workers when conducting range queries on the T-Loc dataset with a selectivity of 0.8\%. \rev{During similarity queries, worker nodes \#1 and \#2 operate for less than 2 seconds, while worker \#10 operates for over 10 seconds. This indicates that nodes \#1 and \#2 are mostly idle while waiting, while node \#10 remains constantly active throughout the query process. However, considering the total workload for all these ten workers only requires one worker node to operate for approximately 50 seconds, achieving a balanced workload where each node operates for 5 seconds could reduce the total query time from over 10 seconds to 5 seconds. Hence, our objective is to develop an effective distributed index for similarity search in metric spaces with a balanced workload.} Nevertheless, three challenges must be addressed to achieve this objective.


\textit{Challenge I: How to efficiently partition the metric space locally?} In distributed environments, effective partitioning of objects into primary nodes and workers is crucial~\cite{dase/HaS22}. Primary nodes control the overall data distribution, while workers handle local \rev{divisions}. Existing methods for distributed object partitioning can be classified into two categories: homogeneous partitioning~\cite{shang2018dita} and heterogeneous partitioning~\cite{icde/ZhengWZZ0J21}. Homogeneous partitioning distributes similar objects to the same nodes, {providing excellent pruning capabilities but under-utilizing computing resources. For instance,}
when performing hot spot queries for a set of similar objects, only a few nodes are involved in computation.
On the other hand, heterogeneous partitioning evenly distributes similar objects across all \rev{groups}, maximizing computing resources. {However, it lacks efficiency in eliminating unnecessary distance computations by pruning clusters that group similar objects.}
Thus, we present a novel three-stage partitioning strategy. Firstly, we employ homogeneous partitioning, leveraging the global index of the primary node to perceive the overall object distribution and {enhance pruning capabilities}. Secondly, we perform heterogeneous partitioning to evenly distribute objects among workers, {optimizing the utilization of computing resources.}
Lastly, within each heterogeneous subregion in each worker, we apply homogeneous partitioning and use local indexing for efficient internal management of local \rev{groups}.

\textit{Challenge II: How to make full use of computing resources?}
In distributed similarity queries, a common issue arises when similar query tasks are assigned to the same computing node, resulting in load imbalance and hindering the efficiency of parallel queries. The most commonly used solution~\cite{shang2018dita} is to create dual data copies in different workers, resulting in data redundancy and heavy communication overhead when backing up objects.
To tackle these issues, we propose a three-stage similarity search strategy. Firstly, we conduct preliminary query searches using a global index to prune unnecessary partitions, employing pruning and verification techniques to further reduce the search region and query radius. Secondly, we leverage the intermediate index to accurately locate the heterogeneous partitions that need to be queried based on the refined query range. Finally, we allocate query tasks to workers and perform precise queries by utilizing homogeneous partitions with local indexes. Since heterogeneous objects are partitioned among workers, all computing resources actively participate into the calculation for any query.

\textit{Challenge III: How to support efficient distributed similarity search?}
Although the proposed three-stage partitioning method and the novel indexing structure can evenly \rev{divide} objects and fully utilize all computing resources, the challenge remains in improving the efficiency of distributed similarity search while reducing communication overhead. Specifically, during the querying process, we first prune unnecessary object partitions in the primary node and then allocate query tasks to workers for precise distance calculations. However, storing a large number of objects in the primary node leads to a significant increase in query tasks, resulting in high communication cost for task allocation to computing nodes.
This raises the need to control the number of objects stored in the primary node, which may weaken its pruning and validation capabilities. To overcome this, we comprehensively consider data partitioning, index construction, query efficiency, and communication overhead. We propose a cost model for distributed similarity search that optimizes the distribution of objects between the primary node and workers through theoretical analysis. This optimization enhances distributed pruning capabilities while reducing communication overhead, thereby supporting efficient query performance.


In summary, we make key contributions as follows:
\begin{itemize}\setlength{\itemsep}{-\itemsep}
\item{}  \textit{Distributed indexing.} We present DIMS, a \underline{D}istributed \underline{I}ndex for similarity search in \underline{M}etric \underline{S}paces, which supports efficient metric range query and metric \textit{k} nearest neighbour query. 
\item{} \textit{Effective objects partition.} We design a novel distributed indexing structure with a three-stage object partition method to efficiently distribute objects evenly among distributed worker nodes, which captures the characteristics of various data and achieves balanced workloads.
\item{} \textit{Distributed similarity search.} We develop concurrent search methods for our proposed distributed index and three-stage partition to support efficient concurrent similarity search, and leverage filtering and validation techniques to avoid unnecessary distance computation.
\item{} \textit{Cost optimization.} We devise cost-based optimization technique with workload adjustment strategy to strike the balance between communication cost and computation cost.
\item{} \textit{Extensive experiments.} We conduct extensive experimental evaluation on five real datasets. The results demonstrate that DIMS outperforms existing distributed similarity search approaches significantly.
\end{itemize}

The rest of this paper is organized as follows. We provide a review of previous works in Section~\ref{sec:relatedwork} and present the problem statement in Section~\ref{sec:problemformulation}. Subsequently, we introduce the distributed index for metric spaces in Section~\ref{sec:framework} and detail the similarity search process in Section~\ref{sec:search}. Finally, we report comprehensive experimental studies in Section~\ref{sec:exp} and conclude the paper in Section~\ref{sec:conclusion}.

\vspace{-2mm}
\section{Related Work}
\label{sec:relatedwork}
\vspace{-1mm}

\begin{table}\label{tab:symbol}
	\centering
	\renewcommand\arraystretch{1.1}
	\caption{{Symbols and description}}
\vspace{-3mm}
	\small
	\setlength{\tabcolsep}{3pt}
	\begin{tabular}{p{2.2cm}p{6cm}}
		\hline
		\textbf{Notation} & \textbf{Description} \\
		\hline
        $q$, $o$ & A query, an object in a metric space\\
        $O$ & An object set in a metric space\\
        {$N$} & The number of objects  \\

        $d(\cdot,\cdot)$ & A distance metric\\
        $p$ & An object partition\\
        {$N_p$} & The number of partitions  \\
        {$N_p^*$} & The optimized number of partitions  \\
        $\textit{MkNNQ}(\cdot, \cdot)$ & A metric $k$ nearest neighbour query\\	
        $\textit{MRQ}(\cdot, \cdot)$ & A metric range query\\	

		\hline
	\end{tabular}
	\vspace{-4mm}
\end{table}

In this section, we review the existing works on metric indexes and distributed similarity search.

\vspace{-2mm}
\subsection{Metric Indexes}
\vspace{-1mm}
Similarity search in metric spaces has been widely studied. Based on object partition, filtering, and validation, similarity queries can be answered efficiently by existing approaches, which can be classified into three categories~\cite{ChavezPromximity,chen2020indexing}, i.e., \emph{compact partitioning methods}, \emph{pivot-based methods}, and \emph{hybrid methods} that combine the previous two techniques.

Compact partitioning methods divide objects into compact sub-regions, and leverage the region radii to prune unqualified partitions, including List of Clusters~\cite{prl/ChavezN05,spire/ChavezN00}, Generalized Hyperplane Tree~\cite{ipl/Uhlmann91}, Bisector Tree~\cite{tse/KalantariM83}, Spatial Approximation Tree~\cite{is/ChavezLRR16,vldb/Navarro02}, M-tree~\cite{vldb/CiacciaPZ97}, etc.
Different from compact partitioning methods that leverage partition centers for object pruning, pivot-based methods map the metric space into vector spaces with a set of pivots, and prune objects with vector indexing approaches to boost similarity search, such as Linear AESA~\cite{prl/MicoOV94}, MVP Tree~\cite{tods/BozkayaO99,sigmod/BozkayaO97}, Spacing-filling curve and Pivot-based B$^+$-tree~\cite{tkde/ChenGLJC17,icde/ChenGLJC15}, etc.
To further accelerate similarity search, hybrid methods combine compact partitioning with the use of pivots. The Geometric Near-Neighbor Access Tree~\cite{vldb/Brin95} utilizes the generalized hyperplane partition for dividing objects, and employs the pivot-based cut-regions for object filtering. The M-index~\cite{is/NovakBZ11} proposes the iDistance technique~\cite{tods/JagadishOTYZ05} for compacting objects, which is one of the most efficient indexes as  mentioned in the latest survey on metric similarity search methods. 
{However, all the above methods cannot deal with the case when the data cardinality exceeds the storage capacity or processing capacity of a single machine.}


\vspace{-2mm}
\subsection{Distributed Similarity Search}
\vspace{-1mm}
As the demand for higher search performance increases, many recent studies have focused on distributed similarity search, which can be partitioned into distributed implementations of existing metric indexes and distributed indexes for specific metric spaces.

Methods falling into the first category leverage various techniques, such as metric partition or pivot-mapping, to distribute objects, which are then indexed using metric indexes by worker nodes. For example, GHT*~\cite{delos/BatkoGZ04} employs a ball partition strategy to compact objects and utilizes Generalized Hyperplane Tree for worker nodes. Similarly, M-Chord, MT-Chord, and M-index~\cite{is/NovakBZ11,dpd/DoulkeridisVKV09,waim/ZhuSKNY12} apply iDistance~\cite{tods/JagadishOTYZ05} for global object partitioning and employ the B$^+$-tree structure for local indexing. Recently, the Asynchronous Metric Distributed
System~\cite{dase/YangDZCZG19} proposes to partition objects via pivot-mapping into minimum bounding boxes, and utilizes the publish/subscribe communication mode to support asynchronous processing. {On the other hand, DIMA~\cite{pvldb/SunSLDB17} employs a hash map for distributed similarity selection and join operations, but it lacks support for exact similarity search and nearest neighbour queries.}
Nevertheless, existing studies lack a cost model for workload balance, a global and local index structure for effective indexing, and efficient concurrent search approaches.

To address load imbalance and enhance similarity search performance, various distributed indexes have been proposed for specified metric spaces, such as trajectories similarity and time series similarity. For instance, DITA~\cite{shang2018dita} and REPOSE~\cite{icde/ZhengWZZ0J21} use a reference point based trie index to organize trajectory data for local indexing, and devise workload adjustment strategy to eliminate load imbalance for efficient trajectory similarity search. 
DPiSAX~\cite{tkde/YagoubiAMP20} utilizes a sampling-based partitioning table to group time series, which are then distributed into parallel iSAX indexes across worker nodes. \rev{D-HNSW~\cite{bigdataconf/DengYNJC19} extends SOTA graph-based method HNSW~\cite{bigdataconf/DengYNJC19} for distributed environments by partitioning the objects using sampled graph at the primary node and building individual graphs on each worker.}
However, these distributed indexes are developed for specific data types and fail to support similarity search in general metric spaces. To this end, we propose a new distributed index designed to efficiently index metric spaces. Our approach ensures workload balance through a heterogeneous partition method guided by a cost model. Furthermore, we also develop efficient search methods to support concurrent similarity searches with high performance.

\vspace{-1mm}
\section{Problem Formulation}
\label{sec:problemformulation}
\vspace{-1mm}

We proceed to introduce the metric space and the similarity search. Table~\ref{tab:symbol} summarizes frequently used notations. 

A metric space is defined as a tuple $(M, d)$, where $M$ is the domain of objects, and $d$ is a distance metric to quantify the similarity between any pair of objects $(o,q)$ in this space. The distance metric $d$ should satisfy the following conditions: (i) symmetry: $d(q, o)=d(o, q)$; (ii) non-negative: $d(q, o) \geq 0$; (iii) identity: $d(q, o)=0$ iff $q = o$; and (iv) triangle inequality: $d(q, o)\leq d(q, o^\prime) + d(o, o^\prime)$. Manhattan distance, Euclidean distance, and word edit distance are examples of distance metrics that can be used in metric spaces.
Note that, metric space has no requirements for object formulation and can accommodate any data type. Based on the above, we can formally define two types of metric similarity search.

\begin{definition}  
\label{defn:MRQ}
{\bf (Metric Range Query.)} Given an object set $O$, a query object $q$, and a search radius $r$ in a metric space, a metric range query (MRQ) finds the objects in $O$ that are within distance $r$ from $q$, i.e., $MRQ(q, r) = \left\{o\vert \ o \in O \land d(q, o) \leq r\right\}$.
\end{definition}

\begin{definition}
\label{defn:kNNQ}
{\bf (Metric $k$ Nearest Neighbor Query.)} Given an object set $O$, a query object $q$, and an integer $k$ in a metric space, a metric $k$ nearest neighbor query (M\textit{k}NNQ) finds $k$ objects in $O$ that are most similar to $q$, i.e., $MkNNQ(q, k)= \{S\vert \ S \subseteq O \land \vert S\vert = k \land \forall s \in S, o \in O-S$,
$d(q, s) \leq d(q, o)\}$.
\end{definition}

\begin{figure}
\includegraphics[width=0.99\linewidth,height=0.22\textwidth]{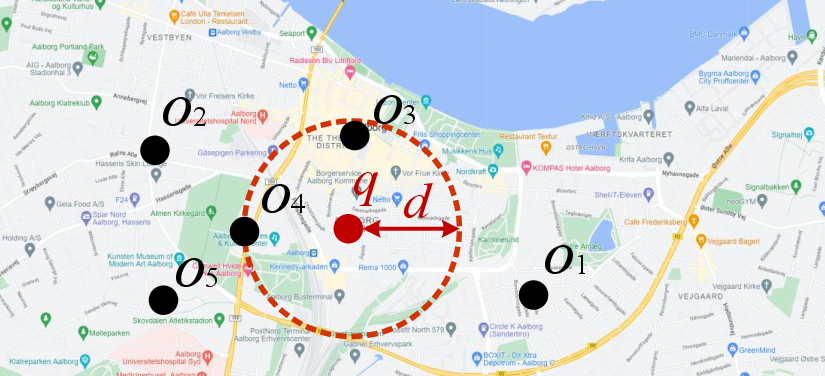}\vspace{-2mm}
  \caption{{Metric space and similarity search}}
  \vspace{-4mm}
  \label{fig:m-examp}
\end{figure}

\textbf{Example I.} Consider a location object set as shown in Fig.~\ref{fig:m-examp}, where $O = \{o_1,o_2,o_3,o_4,o_5\}$, and $L_2$-norm distance is employed to quantify the similarity between objects. Given the query $q$, a metric range query with search radius $r$ ($=d$) finds the objects that locate near the query within the radius $d$, i.e., $MRQ(q, d) = \{o_3, o_4\}$. An example of metric $k$ (= 2) nearest neighbor query finds 2 objects from $O$ with the closest distances to the query, i.e., $MkNNQ(q, 2) = \{o_3,o_4\}$. Note that, if the distance from the query object to its $k$-th nearest neighbour is predetermined, an M\textit{k}NNQ can be solved by means of an MRQ. For instance,  $MkNNQ(q, 2)$ can be answered by $MRQ(q, d)$ if the distance $d=d(q,o_4)$ is given in advance.

\textbf{Example II.} Consider a string dataset $O = \{``00100"$, $``10111", ``01001", ``0110"\}$ and edit distance. Given a query  $q=``10110"$, the answer of $MRQ(q, 1)$ is $\{``10111"$, $``0110"\}$, as they can be changed to $``10110"$ within 1 edit operation including insertion, deletion, or replacement. The metric $k$ ($=3$) nearest neighbour query  $MkNNQ(q, 3)$ retrieves 3 objects that can be modified to $q$ with the least number of operations, yielding $\{``10111", ``0110", ``00100"\}$.

\vspace{-2mm}
\section{Distributed Indexing}
\vspace{-1mm}
\label{sec:framework}
In this section, we first provide an overview of distributed indexes for metric spaces. Following this, we introduce the framework of our proposed method DIMS. Next, we detail the three-stage object partition strategy that integrates both homogeneous and heterogeneous object partitions. Finally, we discuss the implementation of the distributed metric index.

\vspace{-2mm}
\subsection{Overview }
\vspace{-1mm}
Although many distributed approaches have been proposed to accelerate similarity search in metric spaces, their straightforward object partitioning strategies and indexing structures have limitations on search performance. Specifically, existing methods employ a two-layer framework for indexing objects. Initially, they partition the objects with a global index using homogeneous or heterogeneous partition strategies, and then distribute the objects into worker nodes with local indexes. However, both homogeneous and heterogeneous partitioning strategies have drawbacks that affect  workload and computation cost balance. Homogeneous partitioning can lead to unavoidable computing resource waste, whereas heterogeneous partitioning cannot efficiently prune objects, resulting in unnecessary distance computations.
One possible solution is to combine both partition strategies: employing homogeneous partition for all objects in the primary node, followed by heterogeneous partitioning of objects in the worker nodes. To enhance local search performance, homogeneous partitioning with metric indexes can be leveraged for each worker. However, transforming the partition result of each stage through network communication is time-consuming and may not be efficient.

\begin{figure}
\centering  \includegraphics[width=0.99\linewidth,height=0.26\textwidth]{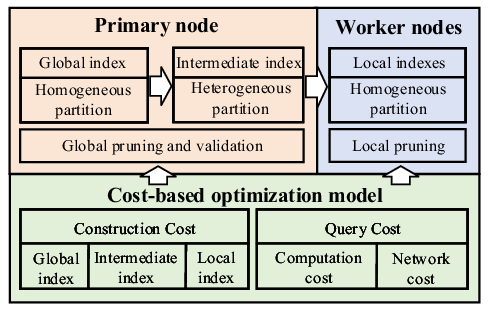}
  \vspace{-4mm}
  \caption{{DIMS framework
  }}
  \vspace{-4mm}
  \label{fig:framew}
\end{figure}

\begin{figure*}
\centering
  \includegraphics[width=0.98\linewidth,height=0.20\linewidth]{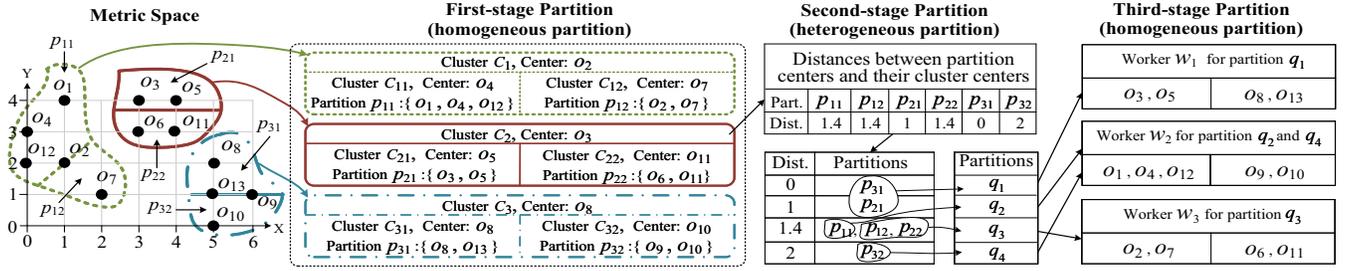}
  \vspace{-3mm}
  \caption{\rev{Illustration of partitioning}}
  \vspace{-3mm}
  \label{fig:parti}
\end{figure*} 

{In Fig.~\ref{fig:framew}, we present three main steps of our DIMS framework. First, we homogeneously partition the objects in the primary node using a global index, and then distribute those partitions heterogeneously through the intermediate index. This enables DIMS to perform global pruning and validation {on clusters that group similar objects.} Next, we allocate the heterogeneous partitions from the intermediate index to workers, and utilize local indexes for efficient object management, achieved by homogeneous partitioning techniques (e.g., clusters). This enables effective local pruning. Finally, we use a cost-based optimization model that considers both construction cost (for global, intermediate, and local indexes) and query cost (including computation cost and network cost) to further improve the performance of DIMS. }

\begin{figure*}
\centering
 \subfigure[Primary node]{
    \label{fig:index-a}
    \includegraphics[width=0.65\linewidth,height=0.5\textwidth]{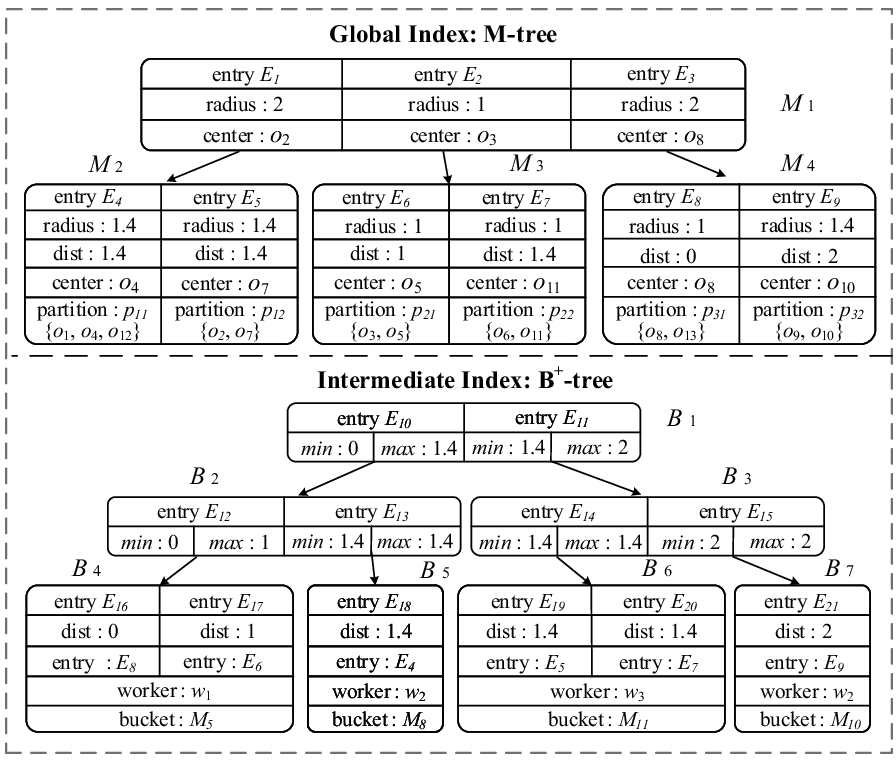}
    \vspace{-0.4cm}
  }
  \subfigure[Worker nodes]{
    \label{fig:index-b}
    \includegraphics[width=0.3\linewidth,height=0.5\textwidth]{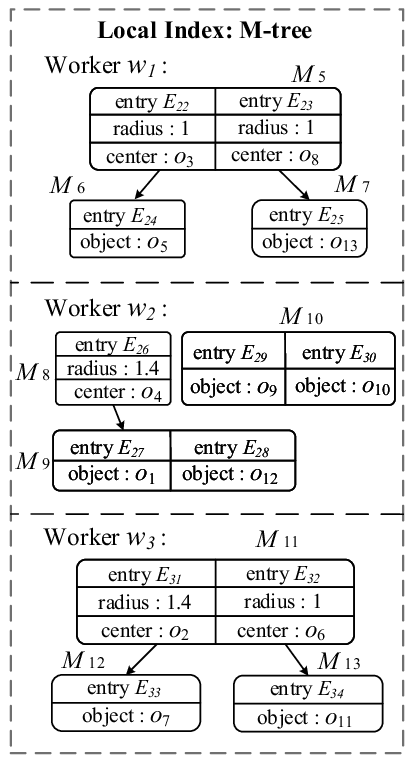}
    \vspace{-0.4cm}
  }
  \vspace{-4mm}
  \caption{\rev{Distributed indexing}}
  \vspace{-5mm}
  \label{fig:index}
\end{figure*}

\vspace{-2mm}
\subsection{Partitioning}
\vspace{-1mm}

In general metric spaces, the absence of a coordinate structure poses challenges to direct object partitioning~\cite{jda/MaoMM12}.
Traditional partitioning methods like grid partitioning, which rely on coordinate information and are commonly used in Euclidean spaces, 
are not applicable in metric spaces lacking a coordinate structure. For example, in text mining applications, grid partitioning is not feasible as there is no inherent coordinate structure in the text data. To overcome this hurdle, we leverage distance estimation and clustering techniques, and propose a three-stage partitioning strategy to ensure even distribution of objects. The number of partitions is guided by a cost model to be detailed in Section~\ref{sec:cost}. To illustrate, we use Fig.~\ref{fig:parti}, considering an object set $O=\{o_1,o_2,...,$ $o_{13}\}$ and adopting $L_2$-norm distance.

\textbf{Step I.} We use clustering techniques to hierarchically divide objects into homogeneous subregions $p_i$, using a set of objects as cluster centers. 
We ensure that each object is assigned to its nearest cluster to group similar objects together.  
{Note that, for simplicity, we utilize the random strategy of M-tree~\cite{vldb/CiacciaPZ97} to form cluster centers, i.e.,  all the centers are selected randomly. }
As depicted in Fig.~\ref{fig:parti}, three center objects ($o_2$, $o_3$, and $o_8$) are first used to divide the objects into three clusters $C_1$, $C_2$, and $C_3$. Each cluster is further partitioned using centers $\{o_4,o_5,o_7,o_8,o_{10},o_{11}\}$, resulting in the partitions $\{p_{11},p_{12},p_{21},p_{22},p_{31},p_{32}\}$.


\textbf{Step II.} Next, we heterogeneously divide objects into sub-regions. 
Firstly, we sort the bottom-level clusters based on the distances from their cluster centers to the centers of their upper-level clusters \rev{(e.g., for $p_{11}$, the distance used for sorting refers to the distance from its center $o_4$ to the center $o_2$ of its parent cluster $C_1$, {i.e., 1.4})}. 
We then merge bottom-level clusters with similar distance values to create new partitions {$q_j$}. 
This approach groups objects with similar distances to their cluster centers into the same partition,
regardless of their dissimilarity.
Thus, each new partition consists of heterogeneous objects.
As illustrated in Fig.~\ref{fig:parti}, both partitions $p_{12}$ and $p_{22}$ are included into the partition $q_3$ because their distances to their corresponding centers are both 1.4, even though these two \rev{groups} are heterogeneous. \rev{Notably, the clusters are partitioned based on their distances and number of objects in this step, ensuring that resulted partitions have clusters of similar distance and contain a similar number of objects.}
Consequently, partitions $q_1$, $q_2$, $q_3$, and $q_4$ all contain approximately 4 objects.

\textbf{Step III.} {Finally, we divide the heterogeneous partition $q_j$ into worker nodes to achieve workload balance, and further homogeneously partition each $q_j$ via clusters for effective management.} As illustrated in Fig.~\ref{fig:parti}, partition $q_3$ is assigned to worker node $w_3$, where objects $o_2$ and $o_{7}$ are placed in the same group as they are similar (i.e., they are grouped together in Step I as part of cluster $C_{12}$), while objects $o_6$ and $o_{11}$ are grouped together.

\vspace{-2mm}
\subsection{Indexing}
\label{sec:indexing}
In DIMS, we propose a three-stage indexing structure that effectively manages metric space objects in a distributed environment. The first stage involves deploying a global index, which perceives the general distribution of all the objects.
The second stage uses an intermediate index to divide objects into heterogeneous partitions, aiming to achieve workload balance among worker nodes. Once the partitions are evenly distributed, the final stage constructs local indexes to facilitate effective internal data management. Fig.~\ref{fig:index} depicts an example of our three-stage indexing structure applied to the metric space shown in Fig.~\ref{fig:parti}. 
Please note that the main focus of this paper is on the development and optimization of the distributed similarity search framework; thus, we leverage the most effective existing indexes to efficiently manage objects in different partition stages. Specifically, because homogeneous partitioning is designed to support efficient pruning, we choose M-tree~\cite{vldb/CiacciaPZ97} as the global and local index for DIMS, as it leverages clusters to group similar objects and uses multiple tree levels to manage clusters for effective objects pruning. Aiming at efficiently managing the distances between objects during heterogeneous partitioning, we apply the B$^{+}$-tree structure. The following sections explain the three-stage indexing structure in detail.

\noindent
\textbf{Global index.} The first stage of DIMS is the global index, which groups objects into homogeneous partitions. 
The global M-tree has two types of tree nodes: non-leaf nodes that store the cluster information, including center objects and cluster radii, and leaf nodes that store object \rev{groups}. Specifically, each leaf entry $E$ in the leaf node maintains the center, the radius, the distance (denoted as \textit{dist}) between its center and the parent entry's center, and the corresponding partition. As shown in Fig.~\ref{fig:index-a}, leaf entry $E_4$ maintains $E_4.center = o_4$, $E_4.radius = 1.4$, $E_4.dist = 1.4$, and $E_4.partition = \{o_1, o_4, o_{12}\}$. Note that, the global index does not store the real objects in each partition but only the partition pointers. Since the primary node has limited storage capacity, we control the size of M-tree by setting partition size boundaries.

Algorithm~\ref{algo:globalindex} presents the detailed steps to build the global M-tree. An example of global index construction is illustrated in Fig.~\ref{fig:index-a}, corresponding to the objects shown in Fig.~\ref{fig:parti}. First, we estimate the size of each partition $N'$ (i.e., the number of objects in each leaf entry of the M-tree) according to the total number of objects $N$ and the number of partitions $N_p$ (line 1). In the example, we set $N'$ to 3 based on the number of centers (i.e., 13) and the number of partitions (i.e., 6), ensuring that each leaf entry is assigned with no more than three centers. Next, we construct the M-tree by assigning each object to its closest cluster (i.e., the closest leaf entry) (lines 3--8). For example, object $o_7$ has the smallest distance to the non-leaf entry center $o_2$ in the root node $M_1$ and thus is assigned to leaf node $M_2$, while $o_7$ is assigned to leaf entry $E_5$ in leaf node $M_2$. Thereafter, the algorithm splits a leaf node when its number of objects exceeds $N'$ (lines 9--10), {which chooses new centers randomly and forms new clusters}.  Finally, we index all the object partitions in the leaf entries $E$ of the global index using the intermediate index (lines 11--12), and return the global M-tree (line 13).

\begin{algorithm}[t]
\small
\SetNlSty{small}{}{:}
\LinesNumbered
\setstretch{0.9}
\DontPrintSemicolon
\caption{Global\_Index}
\label{algo:globalindex}
\KwIn{an object set $O$, the number $N_p$ of partitions }
\KwOut{the global M-tree index}
    $N\gets |O|,N'\gets N/N_p$\;
    \ForEach{ object $o \in O $}
    {
        $T\gets$ the root of M-tree\;
        \While{$T$ is not a leaf node}
        {
            $E\gets \arg \min_{E \in T.entries} d(E.center,o)$ // find the closest cluster to the object $o$\;
            $T\gets$ the tree node pointed by $E$\;
        }
        $E\gets \arg \min_{E \in T.entries} d(E.center,o)$ // find the closest partition of leaf node $T$ to the object $o$\;
        $E.partition\gets E.partition \cup \{o\}$ \;
        \If{$|E.partition|>N'$}
        {	
            split $E$ with M-tree {random} split strategy\;	
        }
    }
    \ForEach{ leaf node entry $E$}{
        Intermediate\_Index($E.partition$)\;
    }
    \KwRet the M-tree\;
\end{algorithm}

\noindent
\textbf{Intermediate index.} 
\rev{The intermediate index divides partitions generated by the global index into homogeneous groups, which are subsequently evenly distributed among worker nodes. Recent research~\cite{icde/ZhengWZZ0J21} highlights the importance of an effective heterogeneous partitioning strategy to ensure partitions contain query results by grouping objects into similar partitions and then sorting them based on their cluster id before partitioning in a round-robin fashion. 
Similarly, our proposed DIMS index first partitions object groups generated by the global M-tree index based on distances between their centers and their parents' centers. These groups are then managed by their distances to their parent nodes using a high-performance index structure B$^+$-tree~\cite{fcsc/JinCLLW24}. 
Finally, the DIMS distributes leaf nodes of the B$^+$-tree in a round-robin fashion among worker nodes, facilitating the scattering of dissimilar objects for efficient processing.}
Notably, both the global M-tree index and the intermediate B$^+$-tree index have similar space consumption, as they share the same total number of leaf entries. Therefore, the intermediate index can also be stored in the primary node. {Given its optimized memory access and fast range query support~\cite{sigmod/ShahvaraniJ16}, we adopt the B$^+$-tree as the intermediate index. Considering that the intermediate index manages a relatively small subset of data groups compared to the entire dataset, we also conducted experiments to compare the efficiency of the B$^+$-tree with a linear scan strategy (i.e., without an intermediate index). The results, presented in Section~\ref{sec:ssp}, demonstrate the superior performance of the intermediate B$^+$-tree index.}


Algorithm~\ref{algo:mediumindex} provides a detailed depiction of the construction of the intermediate B$^+$-tree index, with an example shown in Fig.~\ref{fig:index-a}. 
Consider partition $p_{21}$ in leaf entry $E_6$ of the global index, represented by the similarity distance, which is 1, from its entry center $o_{5}$ to the cluster center $o_3$ of $E_2$ (the parent entry of $E_6$). We first index partition $p_{21}$ using the B$^+$-tree (lines 1--2), placing it at entry $E_{17}$ of the leaf node $B_4$. Note that as $p_{21}$ shares a \rev{relative similar distance (0) as the distance of $p_{31}$ (=2)
compared to the distance of $p_{32}$ (=2).}
Thus, $p_{31}$ is also placed in leaf node $B_4$, even though the objects in $p_{21}$ and those in $p_{31}$ are dissimilar. Additionally, in order to achieve a balanced workload among workers, 
\rev{partitions are assigned to B$^+$-tree nodes based on their similarity distances to parent nodes, while ensuring that each leaf node is allocated an equal number of partitions.
Therefore, some partitions having the same similarity distance are divided into different leaf nodes as the total amount of objects in those partitions might surpass the capacity of a single node. For example, the partitions $p_{11}$ and $p_{12}$ both have distances of 1.4, but they are allocated to leaf nodes $B_{5}$ and $B_{6}$, due to node $B_{6}$ reaching its full capacity of 2 objects.}
Next, all heterogeneous partitions in leaf nodes are assigned to workers (lines 3--4). For instance, leaf node $B_4$ is assigned to worker $w_1$ in bucket $M_5$. {Finally, the intermediate index is returned (line 5).}

\begin{algorithm}[t]
\small
\SetNlSty{small}{}{:}
\LinesNumbered
\setstretch{0.92}
\DontPrintSemicolon
\caption{Intermediate\_Index}
\label{algo:mediumindex}
\KwIn{a partition set $P$}
\KwOut{the intermediate B$^+$-tree index}
    \ForEach{ partition $p \in P $}
    {
        update the B$^+$-tree with $p$\;
    }
    \ForEach{ leaf node $B$ of the B$^+$-tree}{
        Local\_Index($B$)\;
    }
    \KwRet the B$^+$-tree\;
\end{algorithm}
\setlength{\floatsep}{0.1cm}

\noindent
\textbf{Local index.} We proceed to explain how to build a local index for the heterogeneous partitions on workers. This process is similar to existing distributed indexes that build a classical M-tree for local workers. However, there is a difference: a single worker may be allocated with more than one partition in DIMS, meaning that this worker will build multiple M-tree indexes and form an M-forest. Note that, each partition consists of dissimilar objects, which will be homogeneously partitioned by M-tree.

Algorithm~\ref{algo:localindex} presents the local index construction. Consider worker $w_1$ in Fig.~\ref{fig:index-b}, which is allocated with leaf node $B_4$ in the intermediate index. First, the algorithm initializes an empty root node $M_5$ (line 1). Next, it finds two \rev{groups} in $B_4$ (line 2), i.e., partition $p_{31}$ in entry $E_8$ and partition $p_{21}$ in entry $E_4$, and homogeneously clusters all the objects in each partition with constructed M-tree (lines 3--5). Consequently, the objects in the two \rev{groups} are divided into two distinct clusters (i.e., $\{o_3, o_5\}$ and $\{o_8, o_{13}\}$), which have $o_3$ and $o_8$ as their respective centers. Subsequently, the local index is returned (line 6).

\begin{algorithm}[t]
\small
\SetNlSty{small}{}{:}
\LinesNumbered
\setstretch{0.92}
\DontPrintSemicolon
\caption{Local\_Index}
\label{algo:localindex}
\KwIn{a B$^+$-tree leaf nodes $B$}
\KwOut{the local index}
    $T \gets $ a new initialized M-tree\;
    \ForEach{ entry $E \in B$}
    {
    \ForEach{ object $o$ in the partition of entry $E$}
        {            
            insert a new entry $E_o$ for $o$ into $T$\;
            balance $T$ with M-tree {random} split strategy\;
        }
    }
    \KwRet $T$\;
\end{algorithm}

\vspace{-3mm}
\subsection{Complexity Analysis}
\label{sec:complex_a}

\noindent
\textbf{Space consumption.} DIMS consists of three components, i.e., the global M-tree index, the intermediate B$^{+}$-tree index, and the local M-forest. Let $N$ represent the size of objects (i.e., $|{O}|$),
$N_p$ denote the number of partitions, and $N_w$ indicate the number of workers. The storage costs of the global index and the intermediate index are both $O(N_p)$, as the number of leaf entries in both indexes is $O(N_p)$. Regarding the local indexes on each worker, since we randomly allocate the heterogeneous partitions to every local M-tree, the estimated size of each local index is $O(\frac{N}{N_p})$. As $O(\frac{N_p}{N_w})$ partitions are allocated among each worker on average, the space cost of each worker is $O(\frac{N}{N_w})$.

\vspace{0.05cm}
\noindent
\textbf{Time complexity of index construction.} The construction of DIMS requires three steps: (i) building the global M-tree index; (ii) indexing the leaf entries of global index with a intermediate B$^{+}$-tree; and (iii) constructing the local M-forest on each worker. In the first two steps, \rev{the cost of constructing both global and intermediate indexes is $O(N \log N_p)$ as there are $N_p$ leaf entries in global and intermediate indexes.} Next, as there are $O(\frac{N}{N_p})$ objects in each partition and {$\frac{N_p}{N_w}$ partitions in each worker,} the time complexity of constructing each local M-tree index is $O(\frac{N}{N_p}\log \frac{N}{N_p})$, bringing the total index construction cost for each worker to $O(\frac{N}{N_w}\log N)$. 
Additionally, transforming the objects partition from primary node to each worker incurs a communication cost, taking the time complexity of $O(N)$. Therefore, the total construction complexity of DIMS is \rev{$O(N \log N_p+\frac{N}{N_w}\log N+N)$}.

\vspace{-3mm}
\section{Similarity Search}
\label{sec:search}
In this section, we propose efficient distributed algorithms for similarity search in metric spaces using DIMS for metric range query and metric $k$ nearest neighbour query.

\vspace{-3mm}
\subsection{Framework}
In DIMS, similarity queries are processed in four steps. (1) First, the primary node utilizes the global index to compute relevant partitions that contain candidate objects for the given query. (2) Thereafter, it leverages the intermediate index to locate the workers responsible for these partitions and assigns the query tasks to them. (3) Next, each worker generates candidate objects using filtering and validation techniques with local indexes, while the candidates are subsequently verified to obtain local answers. (4) Finally, the primary node collects and verifies the local results, and returns the answers to the similarity query.

\begin{figure}
  \includegraphics[width=0.9\linewidth,height=0.45\linewidth]{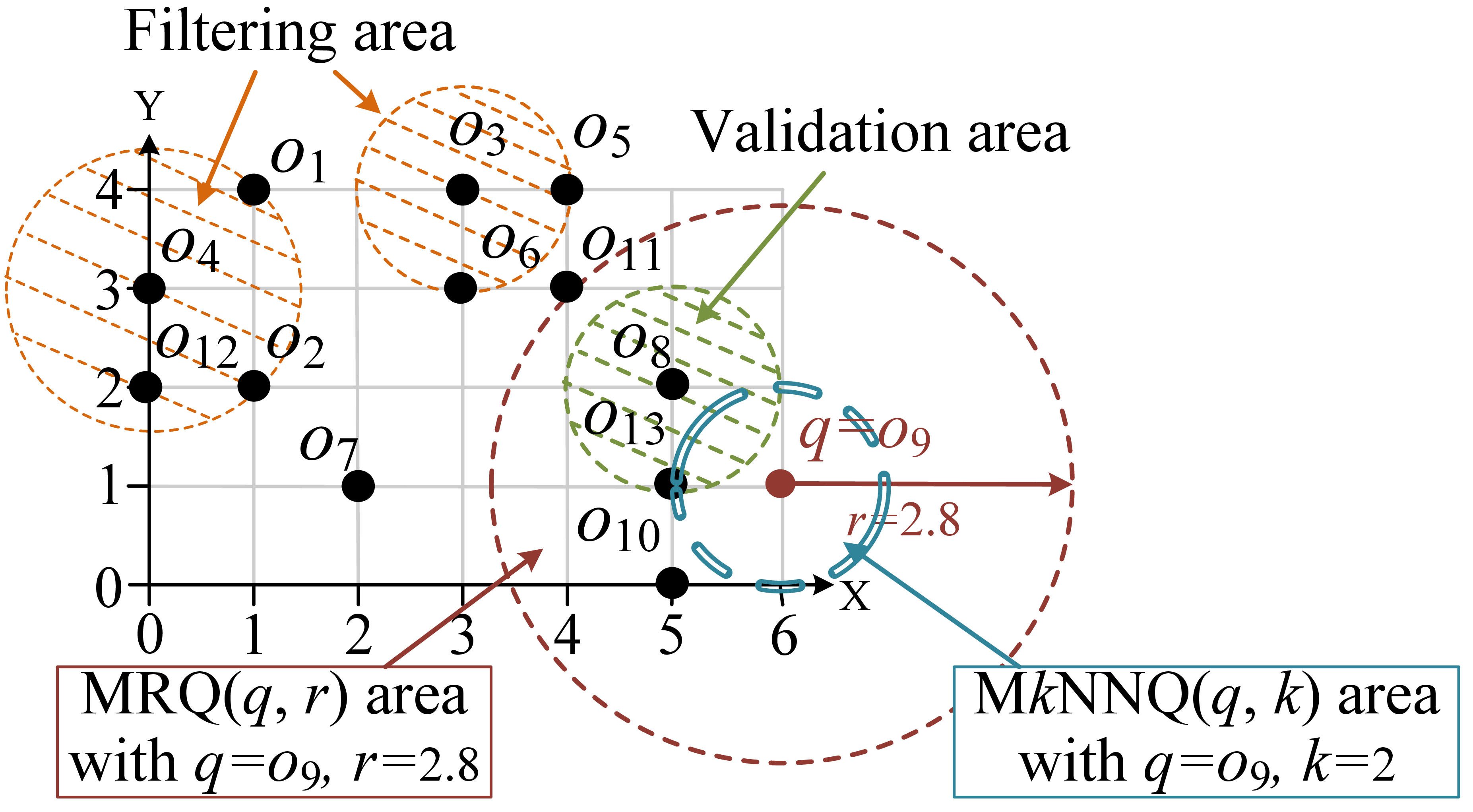}
  \vspace{-3mm}
  \caption{{Illustration of cluster filtering and validation}}
  \label{fig:filter}
\end{figure}

\vspace{-2mm}
\subsection{Metric Range Query}
\label{sec:mrq}

\begin{algorithm}[t]
\small
\SetNlSty{small}{}{:}
\LinesNumbered
\setstretch{0.9}
\DontPrintSemicolon
\caption{{RangeQuery}}
\label{algo:mrq}
\KwIn{a query object $q$, a radius $r$, the DIMS index}
\KwOut{the result set $Ans$ of the range query}
    $Ans\gets \varnothing$\;
    $T\gets$ the root of global index\;
    partition candidates  $Cand\gets$ Range\_Q($T,q,r$)\;
    \ForEach{ partition $p \in Cand$}{
        find the local index $T'$ for $p$ in intermediate index\;
        $Ans\gets Ans$ $\cup$ Range\_Q($T',q,r$)\;
    }
    \KwRet $Ans$\;

    \SetKwFunction{moc}{Range\_Q}
    \textbf{Function} \moc{$T,q,r$}{
    \;
        $Ans\gets \varnothing$\;
        \If{$T$ is a leaf node}
        {
            \If{$T$ belongs to global index}
            {
             \ForEach{ entry $E \in T$, $d(E.center,q)\leq r+E.radius$}{
                    $Ans\gets Ans \cup $ $E.partition$ \;
                }
            }
            \Else
            {
            \ForEach{ entry $E \in T$, $d(E.object,q)\leq r$}{
                    $Ans\gets Ans$ $\cup$ $E.object$ \;
                    }
            }           
        }
        \Else{
            \ForEach{ child node $T' \in T$}{
               \If{$d(T'.center,q)\leq r-T'.radius$}
        	{
                    { // All the objects in $T'$ are verified\;}                    
                    $Ans\gets Ans$ $\cup$ all the leaf entries of $T'$ \;
                }
                \ElseIf{$d(T'.center,q)\leq r+T'.radius$}{
                    { // $T'$ cannot be pruned\;}                    
                    $Ans\gets Ans$ $\cup$ Range$\_$Q($T',q,r$)\;
                }
            }
        }
        \KwRet $Ans$\;
  }
\end{algorithm}

Given a metric object set $O$, a range query with radius $r$ finds the objects in $O$ whose distances to the query object $q$ are bounded by $r$. Consider a metric range query instance in Fig.~\ref{fig:filter} with the same metric space as the example in Fig.~\ref{fig:parti}. The query object $q$ is $o_9$, and the query range $r$ is $2.8$. The answer to the this query $MRQ(q=o_9$, $r=2.8)$ is $\{o_8,o_9,o_{10},o_{11},o_{13}\}$. However, computing the distances between the query object and all the objects to answer range queries can be time-consuming. To accelerate metric range query, we {follow the existing works~\cite{chen2020indexing} and} utilize the triangle inequality to filter and validate objects, in order to avoid unnecessary distance computations.
\begin{lemma} \label{lemmafil}
\rev{Given an object $c$, a query object $q$, and a search radius $r$ in a metric space, an object $o$ can be pruned if $\vert d(o,c) - d(q,c)\vert > r$, while an object $o'$ is validated to be the answer if $d(o',c) + d(q,c) \leq r$.}
\end{lemma}

\begin{lemma} \label{lemmafilc}
\rev{Given a cluster $C$ with center $c$ and radius $r_c$, a query objects $q$, and a search radius $r$ in a metric space, any object $o\in C$ can be pruned for MRQ($q$, $r$) if $d(c,q) > r+r_c$, while any object $o'\in C$ is validated to be the answer if $d(q,c) \leq r-r_c$.}
\end{lemma}

{The proof of Lemmas~\ref{lemmafil}--\ref{lemmafilc} can be directly derived by triangle inequality, and thus, is omitted.} Consider the example shown in Fig.~\ref{fig:filter}, where three clusters are pre-computed, i.e., (1) cluster $C_1=\{o_1,o_2,o_4,o_{12}\}$ with center $o_4$ and radius 1.4; (2) cluster $C_2=\{o_3,o_5,o_6\}$ with center $o_3$ and radius 1; and (3) cluster $C_3=\{o_8,o_{13}\}$ with center $o_8$ and radius 1. According to Lemma~\ref{lemmafilc}, clusters $C_1$ and $C_2$ can be pruned as $d(q,o_4)>1.4+r$ and $d(q,o_3)>1+r$. Meanwhile, objects in $C_3$ are validated to be in the answer to $MRQ(q,r)$ since $d(q,o_8)>r-1$.

{Based on the above lemmas, we design a concurrent index-based search method for metric range query. Specifically, we first search the global index with Lemma~\ref{lemmafilc} to filter and validate candidate clusters, which reduces the total query cost. Next, we utilize the intermediate index to locate the local indexes for the candidates, and assign the query tasks to each associated worker. Note that, as the intermediate index \rev{groups} the objects heterogeneously, the workload of each worker is balanced. Finally, we search the local index in workers, while using Lemma~\ref{lemmafil} to further improve the search performance.}

{Algorithm~\ref{algo:mrq} depicts the detailed steps of the concurrent index-based search method.} It takes a query object $q$, a radius $r$, and the distributed index DIMS, and outputs the result set $Ans$. Initially, the algorithm initializes an empty answer set $Ans$, and conducts the range query in the global index to narrow down the search space and find candidate partitions (lines 1--3). Then, it locates the local index of each candidate partition using the intermediate index, and performs local range queries in the corresponding workers (lines 4--6). Finally, it returns the result set $Ans$ (line 7).

{During range queries, if the current node is a leaf node, DIMS finds all its partitions (for the global index) or objects (for local indexes) that cannot be pruned by Lemma~\ref{lemmafilc} as answers (lines 10--16). For non-leaf nodes, if the entire cluster can be verified, all the leaf entries in this sub-tree are returned as answers (lines 19--21). Otherwise, it iteratively searches the sub-trees that cannot be pruned (lines 22--24). The algorithm finally returns the answer set $Ans$ (line 25).}

\vspace{-2mm}
\subsection{Metric $k$ Nearest Neighbour Query}
\label{sec:mknnq}
\vspace{-1mm}

\begin{algorithm}[t]
\small
\SetNlSty{small}{}{:}
\LinesNumbered
\setstretch{0.9}
\DontPrintSemicolon
\caption{$k$NN\_Query}
\label{algo:knnq}
\KwIn{a query object $q$, an integer $k$, the DIMS index}
\KwOut{the result set $Ans$ of the M$k$NNQ}
    $Ans\gets \varnothing$,$Cand\gets \varnothing$\;
    $D_k\gets \infty$ // the distance of the $k$-th NN object to $q$ \;
    $p\gets$ the nearest partition to $q$ in the global index\;
    $T'\gets$ the local index root for $p$ in intermediate index\;
    $D_t\gets$ $\max_{o\in Local\_\textit{NN}Q(T',q,k,\infty)}$ $d(o,q)$\;
    $Queue\gets \{$ the root $T$ of global index $\}$ // the priority queue to store candidate nodes\;
     \While{$Queue \neq \varnothing$}
     {
            $T\gets Queue.pop()$ \;
             \ForEach{ entry $E\in T$, $d(E.center,q)\leq E.radius+ D_k$ }
             {
                {// $E$ cannot be pruned by Lemma~\ref{lemmaknnfilter}\;}
                \If{$E$ is a leaf entry}
                {
                     $Cand\gets Cand$ $\cup$ $\{E.partition\}$ \;
                    update $D_k$ with $d(E.center,q)+ E.radius$ if necessary // Lemma~\ref{lemmadisfilter}\;
                }
                \Else
                {
                    push the sub-tree node of $E$ and its upper bound distance $d(E.center,q)+ E.radius$ into $Queue$\;
                 }
             }
    }
    \ForEach{ partition $p \in Cand$ }{
        find the local index $T$ for $p$ in intermediate index\;
        $Ans\gets Ans$ $\cup$ Local\_NNQ($T,q,k,min(D_t,D_k$)\;
    }
    keep $k$ objects in $Ans$ that are closest to $q$ \;
    \KwRet $Ans$\;

    \SetKwFunction{moc}{Local$\_$NNQ}
    \textbf{Function} \moc{$T,q,k,D_k$}{
    \;
     $Ans\gets \varnothing$\;
     $Queue\gets \{T\}$ // the priority queue for candidates\;
    \While{$Queue \neq \varnothing$}
     {
        $T\gets Queue.pop()$ \;
         \ForEach{ entry $E\in T$, $d(E.center,q)\leq E.radius+ D_k$ }
         {
            \If{$E$ is a leaf entry}
            {
                update $Ans$ with $\{E.object\}$ \;
                update $D_k$ with $d(E.object,q)$ if necessary\;
            }
            \Else
            {
                push the sub-tree node of $E$ and its upper bound distance $d(E.center,q)+ E.radius$  into $Queue$\;
             }
         }
    }
    keep $k$ objects in $Ans$ that are closest to $q$ \;
    \KwRet $Ans$\;
  }
\end{algorithm}

To answer a metric $k$ nearest neighbor query (M$k$NNQ) in a distributed environment, we can perform M$k$NNQ independently in each worker and verify the local results to obtain the final answer. However, this solution can be expensive since workers are unable to communicate during computation, and each worker cannot leverage the local results of other workers to prune unnecessary objects. Recall that, M$k$NNQ can be answered by MRQ with range $D_k$, where $D_k$ is the distance from the query object to its $k$-th nearest neighbor. Thus, if a tight boundary $D_t$ of $D_k$ is given in advance, each worker can prune objects based on the following lemmas~\cite{chen2020indexing}.

\begin{lemma}
\label{lemmadisfilter}
\vspace{-1mm}
Given a cluster $C$ with center $c$ and radius $r_c$, and a query object $q$, the distance between any object $o\in C$ and $q$ has the upper bound $d(q,c)+r_c$.
\vspace{-1mm}
\end{lemma}


\begin{lemma}
\label{lemmaknnfilter}
\vspace{-1mm}
Given a cluster $C$ with center $c$ and radius $r_c$, a query object $q$, an integer $k$, and a distance $D_t$ that represents the maximum distance from $q$ to $k$ given objects, any object $o\in C$ can be pruned when answering M$k$NNQ($q$, $k$) if $d(q,c) > D_k+r_c$.
\vspace{-1mm}
\end{lemma}

Consider the M$k$NNQ($q=o_9$, $k=2$) for the objects shown in Fig.~\ref{fig:filter}. The answer is $\{o_9,o_{13}\}$, and $D_k$ for the second nearest neighbour $o_{13}$ is 1. During the search, if we know the distance from $q$ to object $o_8$, we can get a tight distance boundary $D_t=d(o_{8},q)+1=2.4$ using Lemma~\ref{lemmadisfilter}. This is because there are two objects in the cluster with  center $o_8$ (i.e., $o_8$ and $o_{13}$) and radius 1. Thus, the maximum distance from any object in this cluster to the query is $d(o_{8},q)+1$. According to Lemma~\ref{lemmaknnfilter}, objects can be pruned if they are not contained in the range query MRQ($q$, 2.4), such as the clusters with the centers $o_3$ or $o_4$.

Based on the above findings, we develop an efficient search method for DIMS to answer M$k$NNQ. First, we locate the nearest partition to the query in the global index, and then use the intermediate index to find the corresponding local buckets. Next, we perform the M$k$NNQ in the local index to obtain a tight distance boundary of $D_k$ and simultaneously conduct a general M$k$NNQ in the global index to obtain candidate partitions. Finally, each worker computes local nearest neighbour results in candidate \rev{groups} with the distance boundary, which are then collected by the primary node to verify the real answers to the given M$k$NNQ.

Algorithm~\ref{algo:knnq} depicts the detailed  M$k$NNQ search process. It takes as inputs a query object $q$, an integer $k$, and the distributed metric index DIMS, and outputs the result set $Ans$. First, the algorithm initializes the answer set $Ans$ and the partition candidate set $Cand$ to empty, and set the distance $D_k$ of the $k$-th NN object to $q$ as infinity. Next, it finds the closest partition $p$ to $q$ in the global index, and obtains a tight distance boundary $D_t$ of $D_k$ by searching the objects in $p$ via the corresponding local index (lines 3--5). Meanwhile, a global search is also conducted to obtain all the candidate partitions. Specifically, the algorithm first initializes the priority queue $Queue$ to the root node of global index (line 6), and then performs a while-loop to find all candidates until the priority queue is empty (lines 7--15). For each entry $E\in Queue$ that cannot be pruned by Lemma~\ref{lemmaknnfilter}, the algorithm incorporates the addition of $E$ to the partition candidate list $Cand$, and tries to update $D_k$, if $T$ is a leaf node (lines 12--13), or pushes $E$ into the priority queue otherwise (line 15). Finally, the algorithm searches the local indexes of all the candidate partitions with the distance bound of the minimum value between $D_t$ and $D_k$, verifies local results from each worker, and returns answer set $Ans$ for M\textit{k}NNQ (lines 16--20). In lines 21--33, we define the nearest neighbour query function for local indexes. Similar to the global query process, it leverages a priority queue to verify entities that cannot be directly pruned by  Lemma~\ref{lemmaknnfilter}, which are then added to the answer list (for leaf entries, lines 28--29), or pushed into queue for further verification (for non-leaf node, line 31). {Lastly, the algorithm returns the local result $Ans$ that keeps the $k$ closest objects to the given query $q$ (lines 32--33).}

\vspace{-2mm}
\subsection{Cost Model}
\label{sec:cost}
In the following, we present a cost model for {MRQ} and {M$k$NNQ} that considers data partitioning, query efficiency, and communication overhead to optimize the object distribution in primary node and workers for better search performance. Firstly, we describe the cost model for {MRQ}.

\noindent
\textbf{Computation cost.} The computation cost for {MRQ} includes global index query cost, intermediate index query cost, and local search cost. We begin with by examining the probability that a partition (object) cannot be pruned by the global (local) index. Since objects are partitioned into different clusters by global and local indexes, we can consider the distances between the query $q$ and the centers of clusters as random variables $X_1, X_2,...,X_{nc}$, where $nc$ is the number of clusters used for pruning. According to Lemma~\ref{lemmafil}, a cluster $C_i$ with center $c_i$ cannot be pruned if $|d(c_i,c)-d(c,q)|\leq r$, making the probability of
\begin{equation}
\vspace{-0.5mm}
P_r(c_i\ is\ not\ pruned)=P_r(|X-Y|\leq r).
\vspace{-0.5mm}
\end{equation}
$Y$ denotes the distance distribution of $d(c,q)$. Since $q$ can also be regarded as a random object, $Y$ follows the identical distribution of $X$. As there are $nc$ clusters and an object cannot be excluded only if it cannot be pruned by all clusters, the probability becomes
\begin{equation}
\vspace{-0.5mm}
P_r(c_i\ is\ not\ pruned)=P_r(|X-Y|\leq r)^{nc}.
\vspace{-0.5mm}
\end{equation}

Since $X$ and $Y$ are two independent identically distributed random variables with variance $\sigma^2$, the mean and variance of $X-Y$ are 0 and $2\sigma^2$, respectively. Using Chebyschev’s inequality, we have
\begin{equation}
\vspace{-0.5mm}
P_r(|X-Y|\leq r)^{nc}\geq(1-\frac{2\sigma^2}{r^2})^{nc}.
\vspace{-0.5mm}
\end{equation}

{Based on the above and $nc$ equals to the index height of $\log N_P$}, we can estimate the lower bound size of partition candidates of global index as $N_p(1-\frac{2\sigma^2}{r^2})^{\log_{N_p}}$. Therefore, the computation cost of primary node is $O(N_p\log_{N_p}(1-\frac{2\sigma^2}{r^2})^{\log_{N_p}})$ (including retrieving candidate partitions in the global index and locating them in the intermediate index). Similarly, as each local index stores $O(\frac{N}{N_p})$ objects, the computation cost of local index is $\frac{N}{N_p}(1-\frac{2\sigma^2}{r^2})^{\log\frac{N}{N_p}}$, while each worker is assigned with $\frac{N_p}{N_w}$ local indexes.

\noindent
\textbf{Communication overhead.} The network transmission cost is proportional to the number of partitions that cannot be pruned by the intermediate index. Let $T_c$ be the communication cost to transfer an entry from the primary node to a worker. The communication overhead can be estimated as $O(N_p(1-\frac{2\sigma^2}{r^2})^{\log_{N_p}}\cdot T_c)$.

As {M$k$NNQ} can be answered by {MRQ}, we can estimate the lower bound of the number of partitions (objects) that cannot be excluded during an {M$k$NNQ} in a similar manner. However, as {M$k$NNQ} incurs an extra query for the nearest cluster to obtain a distance boundary $D_k$, it incurs an additional cost of $O(\log N_p+T_c+\frac{N}{N_p}(1-\frac{2\sigma^2}{r^2})^{\log\frac{N}{N_p}})$. This is much smaller than the query cost in each worker. Hence, the {M$k$NNQ} process has the same cost as {MRQ}.

\noindent
\textbf{Optimization.} {The total time cost of an {MRQ} or {M$k$NNQ} consists of the computation cost and the network overhead. We have previously analyzed their lower bound. By computing the first derivative of total cost function and locating its extremum, we can simplify the expression by omitting constant factors. This leads us to the optimization condition for the total search cost, which is represented by the number $N_p$ of partitions:}
\begin{equation}
\label{equaopt}
\vspace{-0.2mm}
N_p(1+\log N_p\ln(\lambda m)+T_c\ln(\lambda m))\lambda^{2\log N_p}=\frac{N}{N_w}\lambda^{\log N} \ln \lambda,
\vspace{-0.2mm}
\end{equation}
where $\lambda$ represents $1-\frac{2\sigma^2}{r^2}$, and $m$ denotes the number of entries in each M-tree node. Note that,  the left side of Equation~\ref{equaopt} monotonically increases with the growth of $N_p$, while the right side remains constant. This indicates that as $N_p$ increases, the total cost initially decreases and then increases. Thus, the optimized number $N_p^*$ of partitions can be solved efficiently using binary search. {We have verified this in our experiments in Section~\ref{sec:ssp}, where we vary the number $N_p$ of partitions from \rev{50} to \rev{1600}.}

\noindent
\textbf{Discussions.} {Based on the analysis above, our proposed distributed metric index DIMS offers significant efficiency improvements for several reasons. Firstly, we develop a three-stage heterogeneous partitioning technique accompanied by an  indexing structure that evenly distributes objects. This captures the characteristics of all objects and supports efficient local workload balancing and locality-aware data management. 
Secondly, DIMS leverages a global index for objects pruning and verification to avoid unnecessary distance computations. It employs an intermediate index to locate queried heterogeneous object \rev{groups}, ensuring workload balance for various query types (e.g., hot spot query), and utilizes local indexes for efficient retrieval of local results. This enables DIMS to effectively utilize computing resources. Additionally, we develop a cost-based optimization model that balances communication and computation costs. This model considers aspects such as data distribution, index construction, query efficiency, and communication overhead.  It allows DIMS to build a distributed metric index with an optimized structure, thereby enhancing search performance.

\vspace{-2mm}
\section{Experiments}
\label{sec:exp}
In this section, we conduct empirical experiments to evaluate the performance of our proposed method DIMS, including the construction and {update} costs, the similarity search performance, and the scalability.

\vspace{-1mm}
\subsection{Experimental Settings}\label{sec:exp-sett}

\noindent
\textbf{Datasets.} We use \rev{five} real-life datasets in our study: (i) \textit{Words}~\cite{urlmoby} that contains proper nouns, acronyms, and compound words sourced from the Moby project, where edit distance is employed as the metric; (ii) \textit{T-Loc}~\cite{pvldb/GhoshACHSL18} that contains geographical locations of  10$M$ Twitter-users, using the $L_{2}$-norm distance as the distance metric; (iii) \textit{Vector}~\cite{bilbao2018automatic}
that includes 100,000 word embeddings of dimension 300 trained on the Spanish Billion Words Corpus, using \rev{$L_{2}$-norm distance} to measure the similarity between words; (iv) \textit{Color}~\cite{urlflickr} that contains standard MPEG-7 image features extracted from \textit{{Flickr}}, where the similarity between features is measured by the $L_{1}$-norm distance; \rev{and (v) \textit{DEEP}~\cite{corr/abs-2205-03763} that consists of billion-scale features generated by the last fully-connected layer of the GoogLeNet model, using $L_{2}$-norm distance as the distance metric.}
We have summarized the datasets in Table~\ref{tab:datasets}, where Dimen. refers to dimensionality. 

\begin{table}
\caption{Statistics of the datasets used}
\label{tab:datasets}
\small
\vspace{-0.3cm}
\begin{tabular}{|p{1.4cm}<{\centering}|p{1.7cm}<{\centering}|p{1.4cm}<{\centering}|p{2.6cm}<{\centering}|}
\hline
\textbf{Dataset} & \textbf{Cardinality} & \textbf{Dimen.} & \textbf{Distance Metric}\\
\hline
\textit{Words} & 611,756 & 1$\sim$34  & \textit{Edit Distance}   \\ \hline
\textit{T-Loc} & 10,000,000 & 2  & $L_{2}$-norm \\ \hline
\textit{Vector} & 100,000 & 300  & \rev{$L_{2}$-norm} \\ \hline
\textit{Color} & 1,000,000 & 282  & $L_{1}$-norm  \\ \hline
\rev{\textit{Deep}} & \rev{50,000,000} & \rev{96}  & \rev{$L_{2}$-norm}  \\ \hline
\end{tabular}
\vspace{-0.1cm}
\end{table}

\begin{table}
\caption{{Evaluation parameters in our experiments}}
\small
\vspace{-0.3cm}
\renewcommand\arraystretch{1}
\label{tab:quepara}
\setlength{\tabcolsep}{1.05mm}{
\begin{tabular}{| p{4.2cm}<{\centering}  | p{4.2cm}<{\centering} |} \hline
\textbf{Parameter} & \textbf{Value} \\ \hline
Search radius \textit{r} (\%) & 0.1, 0.2, 0.4,  \textbf{0.8}, 1.6, 3.2   \\ \hline \rule{0pt}{8pt}
Integer \textit{k} & 1, 2, 4, \textbf{8}, 16, 32\\ \hline  
\rev{Node fanout}& \rev{5, 10, \textbf{20}, 40, 80} \\ \hline 
\rev{Tuning parameter $N_p$}& \rev{50, 100, \textbf{200}, 400, 800, 1600} \\ \hline \rule{0pt}{8pt}
{Number $N_w$ of workers} & 2, 4, 6, 8, {\textbf{10}}   \\ \hline \rule{0pt}{8pt}
\textit{Cardinality} (\%) & 20, 40, 60, 80, \textbf{100} \\ \hline
\end{tabular}
}
\end{table}

\vspace{0.08cm}
\noindent
\textbf{Baselines.}
To verify the performance of our proposed method DIMS, we compare it against (i) two existing distributed approaches for similarity search: the M-index~\cite{waim/ZhuSKNY12} and AMDS~\cite{dase/YangDZCZG19}; \rev{(ii) the single machine method M-tree~\cite{vldb/CiacciaPZ97}; and (iii) the distributed approximate nearest neighbour search method D-HNSW that implements the graph method HNSW~\cite{pami/MalkovY20} on Spark, which has shown superior search performance compared to other approximate methods such as quantization-based methods in recent empirical studies~\cite{tkde/LiZSWLZL20,pvldb/WangXY021}}, using the distributed framework~\cite{bigdataconf/DengYNJC19}. {The M-index family, including  variants like M-chord and MT-chord, has been proven to outperform other methods 
(excluding AMDS) in previous studies. Notably, AMDS has not been directly compared with the M-index family in previous studies. To further demonstrate the effectiveness of DIMS, we also include \rev{three} baselines derived from DIMS: \rev{(i) DIMS with random strategy to heterogeneously partition the nodes in the global index to workers (DIMS-R), i.e., with no intermediate index;} (ii) DIMS with no local homogeneous partition strategy (DIMS-NL);  and (iii) DIMS with pivot-based index (MVPT~\cite{tods/BozkayaO99}, which is the most efficient in-memory pivot-based metric index according to the most recent metric index survey~\cite{chen2020indexing}) for workers (DIMS-P).
All experiments were conducted on a cluster of 11 nodes, each equipped with two 12-core processors (E5-2620 v3 2.40 GHz), 128GB RAM, Ubuntu 14.04.3, Hadoop 2.6.5, and Spark 2.2.0. \rev{Notably, we extend Spark to create the indexes over RDDs, as 
Spark is a fundamental computation framework for big data analytics in real applications~\cite{sigmod/Zhang0SYJM19}.}
Meanwhile, we utilize the bulk-loading method~\cite{ciaccia1998bulk} for efficiently constructing DIMS. For both M-index and AMDS, we set the number of partitions according to the default parameter settings mentioned in the corresponding paper.
The source codes of the implemented algorithms were publicly available~\cite{urlopenc}.

\begin{table}[t]
\small
\caption{{Construction costs and storage sizes}}
\label{tab:construction_cost}
\vspace{-0.2cm}
\renewcommand\arraystretch{1.05}
\setlength{\tabcolsep}{1.12mm}{
\begin{tabular}{|p{1.2cm}<{\centering} |p{1.4cm}<{\centering} |p{0.9cm}<{\centering} |p{0.9cm}<{\centering} |p{0.9cm}<{\centering}|p{0.9cm}<{\centering}|p{0.9cm}<{\centering}|} \hline

 \multicolumn{2}{|c|}{Datasets} & \textit{Words} & \textit{T-Loc} &\textit{Vector} &\textit{Color}&\textit{{DEEP}}\\ \hline 
\specialrule{0.05em}{0pt}{0pt}
\multirow{8}{*}{\makecell[c]{Time \\ (s)}} & {M-index} &{47.3} &{30.8} &{{46.6}} &{41.2} &{{--}} \\ \cline{2-7}
 & {AMDS} & {56.7 } &{ 40.8 } &{ {48.4}} &{ 44.5}&{{3182}} \\ \cline{2-7}
&   {{M-tree}} &{{175.4}} &{ {117.1} } &{{ 189.4}} &{ {170.4}} &{{--}}\\ \cline{2-7}
&{{D-HNSW}} &{{304.0} } &{ {208.4} } &{ {351.1} } &{ {277.0}} &{{--}}\\ \cline{2-7}
&{{DIMS-R}}  &{{33.0} } &{ {22.0} } &{ {31.6} } &{ {30.7}} &{{2573}}\\ \cline{2-7}
&{DIMS-NL}  &{32.2 } &{ 21.0 } &{ {30.9 }} &{ 29.5} &{{2065}}\\ \cline{2-7}
& {DIMS-P}  &{32.9 } &{ 21.6 } &{ {31.8} } &{ 30.2} &{{2617}}\\ \cline{2-7}
& {DIMS}  &{33.3 } &{ 21.8 } &{ {38.1} } &{ 31.2} &{{2658}}\\ \hline
\specialrule{0.05em}{0pt}{0pt}
 \multirow{8}{*}{\makecell[c]{Storage\\ (MB)}}& {M-index} & {70 } & { 973 } & {{457} } & { 44526} & {{--}}\\ \cline{2-7}
 & {AMDS} & {73 } & { 1026 } & { {481} } & { 47091}& {{169477}} \\ \cline{2-7}
 & {{M-tree}} & { {61} } & {  {933}} & {  {430} } & {  {42281}} & {{--}}\\ \cline{2-7} 
& {{D-HNSW}} & { {79} } & {  {981}} & {  {490} } & {  {45889}} & {{--}}\\ \cline{2-7}
& {DIMS-R} & { {70} } & {  {937}} & {  {472} } & {  {44543}} & {{161497}}\\ \cline{2-7}
 & {DIMS-NL}  & {68 } & { 955 } & { {450} } & { 44132} & {{157095}}\\ \cline{2-7}
 & {DIMS-P} & {69 } & { 970 } & { {462} } & { 44862} & {{160908}}\\ \cline{2-7}
 & {DIMS}  & {70 } & { 981 } & { {466} } & { 45397} & {{162406}}\\ \hline

 \end{tabular}
}
\end{table}


\begin{table}[t]
\small
\caption{{Effect of the worker number $N_w$ on DIMS's construction} }
\label{tab:update_cost-wk}
\vspace{-0.2cm}
\renewcommand\arraystretch{1.05}
\setlength{\tabcolsep}{1.12mm}{
\begin{tabular}{|p{0.8cm}<{\centering} |p{1.9cm}<{\centering} |p{0.9cm}<{\centering} |p{0.9cm}<{\centering} |p{0.9cm}<{\centering}|p{0.9cm}<{\centering}|p{0.9cm}<{\centering}|} \hline
 \multicolumn{2}{|c|}{Number of workers} & 2 & 4 & 6 & 8 & 10\\ \hline
 \multirow{2}{*}{\textit{T-Loc}} & Time (s) & 87.2 & 61.8 & 34.6& 26.9& 21.8 \\ \cline{2-7} 
 & Storage (MB) & 968 & 983 & 968 & 975 & 981 \\ \hline  
 \multirow{2}{*}{\textit{Color}} & Time (s) & 125.1 & 88.2 & 48.6 & 38.8 & 30.9 \\ \cline{2-7} 
 & Storage (GB) & 44.7 & 45.2 & 44.3 & 45.2 &45.4 \\ \hline 

\end{tabular}
}
\end{table}

\begin{table}[t]
\small
\caption{{Effect of the partition number $N_p$ on DIMS's construction}}
\label{tab:update_cost-np}
\vspace{-0.2cm}
\renewcommand\arraystretch{1.05}
\setlength{\tabcolsep}{1.4mm}{
\begin{tabular}{|p{1.5cm}<{\centering} |p{1.8cm}<{\centering} |p{0.95cm}<{\centering} |p{0.95cm}<{\centering} |p{0.95cm}<{\centering}|p{0.95cm}<{\centering}|} \hline
 \multicolumn{2}{|c|}{Number of partitions} & \textit{Words} & \textit{T-Loc} &\textit{Vector} &\textit{Color}\\ \hline \rule{0pt}{8pt}
  \multirow{2}{*}{{50}} & Time (s) &28.6 & 18.8 & 34.6 & 25.2 \\ \cline{2-6} \rule{0pt}{8pt}
 & Storage(MB) & 68 & 961 & 457 & 44780 \\ \hline  \rule{0pt}{8pt}
 \multirow{2}{*}{{100}} & Time (s) & 31.6 & 21.3 & 39.7 & 28.4 \\ \cline{2-6} \rule{0pt}{8pt}
 & Storage(MB) & 68 & 967 & 460 & 44984 \\ \hline 
 \multirow{2}{*}{{200}} & Time (s) & 33.1 & 21.7 & 37.8 & 30.7 \\ \cline{2-6} \rule{0pt}{8pt}
 & Storage(MB) & 70 & 981 & 466 & 45394 \\ \hline 
 \multirow{2}{*}{{400}} & Time (s) & 35.4 & 22.7 & 44.2 & 31.9 \\ \cline{2-6} \rule{0pt}{8pt}
 & Storage(MB) & 70 & 988 & 470 & 45853 \\ \hline 
 \multirow{2}{*}{{800}} & Time (s) & 41.9 & 27.6 & 51.4 & 37.6 \\ \cline{2-6} \rule{0pt}{8pt}
 & Storage(MB) & 72 & 1011 & 481 & 46871 \\ \hline 
 \multirow{2}{*}{{1600}} & Time (s) & 48.9 & 32.5 & 60.9 & 43.2 \\ \cline{2-6} \rule{0pt}{8pt}
 & Storage(MB) &74 & 1027 & 490 & 47830 \\ \hline  
\end{tabular}
}
\end{table}

\vspace{0.18cm}
\noindent
\textbf{Parameters and Performance Metrics.} In this study, we evaluate the performance of our proposed method DIMS and compare it with its competitors by analyzing the impact of various parameters. Specifically, we vary the following parameters: the search radius $r$, the integer $k$, \rev{the node fanout of M-tree and B$^+$-tree,} the number $N_p$ of partitions, the number $N_w$ of workers, and the cardinality of the dataset. The parameter \textit{r} is used for MRQ, \textit{k} is used for M\textit{k}NNQ, and $N_p$ is the number of partitions that controls the object distributions in global and local indexes, as defined in Section~\ref{sec:indexing}. \rev{For each dataset, we calculate the theoretical optimal value of $N_p^*$ using Equation~\ref{equaopt} with the default search radius ($=0.8$\%). We find that $N_p*$ is approximately $200$. Since specific search radius values are not provided in practical applications, we set $N_p*$ to $200$ in this paper.}
Table~\ref{tab:quepara} lists the key parameters and their detailed values, where the default values are highlighted in bold.
To evaluate the performance of DIMS, we measure several metrics, including index construction {and update} time, running time, workload variance, and throughput. The index construction time includes both the time for global partition (indexing) cost and the construction cost for local indexes.
For each measurement, we report the average performance based on $100$ random queries.

\begin{table}[t]
\small
\caption{{Effect of the update operation}}
\label{tab:update_cost}
\vspace{-0.2cm}
\renewcommand\arraystretch{1.05}
\setlength{\tabcolsep}{1.4mm}{
\begin{tabular}{|p{2.5cm}<{\centering} |p{1.2cm}<{\centering} |p{0.85cm}<{\centering} |p{0.85cm}<{\centering} |p{0.85cm}<{\centering}|p{0.85cm}<{\centering}|} \hline 
 \multicolumn{2}{|c|}{Datasets} & \textit{Words} & \textit{T-Loc} &\textit{Vector} &\textit{Color}\\ \hline 
 \multicolumn{2}{|c|}{Averaged update cost (ms)}  &14.6&18.4&8.7&11.1\\ \hline \rule{0pt}{8pt}
 \multirow{2}{*}{\makecell{ Query \\ Performance}} & Before(s) &1.52&1.88&0.88&1.16\\ \cline{2-6}\rule{0pt}{8pt}
 & After(s) &1.54&1.90&0.89&1.17\\ \hline 
\end{tabular}
}
\end{table}

\begin{table}[t]
\small
\caption{\rev{Effect of the insertions on MRQ performance}}
\label{tab:update_newcost}
\vspace{-0.2cm}
\renewcommand\arraystretch{1.05}
\setlength{\tabcolsep}{1.4mm}{
\begin{tabular}{|p{1.2cm}<{\centering} |p{1.2cm}<{\centering} |p{0.85cm}<{\centering} |p{0.85cm}<{\centering} |p{0.85cm}<{\centering}|p{0.85cm}<{\centering}|p{0.85cm}<{\centering}|} \hline 
 \multicolumn{2}{|c|}{Inserted objects} & {2\%} & {4\%} & {6\%} & {8\%} & {10\%} \\ \hline 
  \multirow{3}{*}{\makecell{ {\textit{Vector}}}} & M-index  & {1.18} & {1.20} & {1.22} & {1.25} & {1.28}\\ \cline{2-7}
 & AMDS  & {1.41} & {1.43} & {1.45} & {1.49} & {1.52}\\ \cline{2-7}
 & DIMS  & {0.86} & {0.87} & {0.89} & {0.91} & {0.97}\\ \hline 
 \multirow{3}{*}{\makecell{ \textit{Color}}} & M-index &1.83&1.86&1.88&1.94&2.02\\ \cline{2-7}
 & AMDS &1.06&2.10&2.14&2.24&2.30\\ \cline{2-7}
 & DIMS &1.12&1.16&1.18&1.20&1.26\\ \hline 
\end{tabular}
}
\end{table}

\vspace{-2mm}
\subsection{Construction {and Update} Performance}
\vspace{-1mm}

First, we compare the construction cost of DIMS with state-of-the-art competitors, using running time and storage size as the performance metrics. The results are presented in Table~\ref{tab:construction_cost}. 
The results demonstrate that, {with similar storage consumption}, DIMS incurs lower construction cost than existing distributed methods including M-index~\cite{waim/ZhuSKNY12}, AMDS~\cite{dase/YangDZCZG19}, \rev{and D-HNSW}. This is due to our more effective object partition strategy and index structure. Specifically, M-index applies iDistance~\cite{tods/JagadishOTYZ05} for objects partitioning while AMDS  leverages the pivots for partitioning. However, both methods have to compute the distances between each pivot and all the objects, incurring high computation costs. In addition, AMDS is a three-layer structure, which incurs high communication cost. In contrast, we leverage cluster-based partition with M-tree to divide objects, which reduces the computation cost. Moreover, we construct both the global index and the intermediate index in the primary node, which improves the communication efficiency. \rev{Notably, when supporting large dataset \textit{DEEP}, M-tree faces memory issue while M-index fails to compute i-Distance within the master node. Besides, for the reason that D-HNSW builds only one graph in each worker, while the size of RDDs on worker nodes is limited by Spark, D-HNSW fails to support large datasets such as \textit{DEEP}. Therefore, the corresponding results are unreported.}

{Next, we vary the number of workers and the number of partitions during the construction of DIMS to demonstrate their respective impacts on construction. The results are summarized in Tables~\ref{tab:update_cost-wk} and~\ref{tab:update_cost-np}, \rev{revealing that construction time increases with fewer workers  or higher partition counts. Conversely, changes in storage requirements are almost negligible. This finding aligns with our analysis presented in Sec.~\ref{sec:complex_a}.}
}

In addition, we investigate the impact of update operations on DIMS. Notably, since all indexes utilized in DIMS (i.e., M-tree and B$^{+}$-tree) are dynamic metric indexes, DIMS supports object insertions and deletions.
Thus, we compare the efficiency of similarity range queries before and after performing $1,000$ update operations to evaluate their effect. Each update operation involves randomly removing an object from DIMS and then reinserting it. \rev{Meanwhile, we evaluate the similarity search performance following the insertion of new objects (increasing from 2\% to 10\% of the dataset cardinality). 
The results, presented in Tables~\ref{tab:update_cost} and~\ref{tab:update_newcost}, indicate that DIMS supports efficient object update. Moreover, its search performance scales linearly with the expanding dataset size, indicating consistent search efficiency despite update operations.}

\begin{figure}[t]
\begin{center}
\subfigtopskip=-10pt
\subfigcapskip=-4pt
\vspace{0.1cm}
\includegraphics[height=0.78cm]{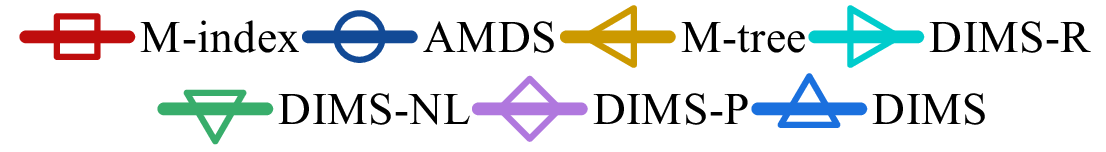}
\vspace{0.01cm}

\subfigure[\textit{Words}]{
  \includegraphics[width=0.225\textwidth,height=0.14\textwidth]{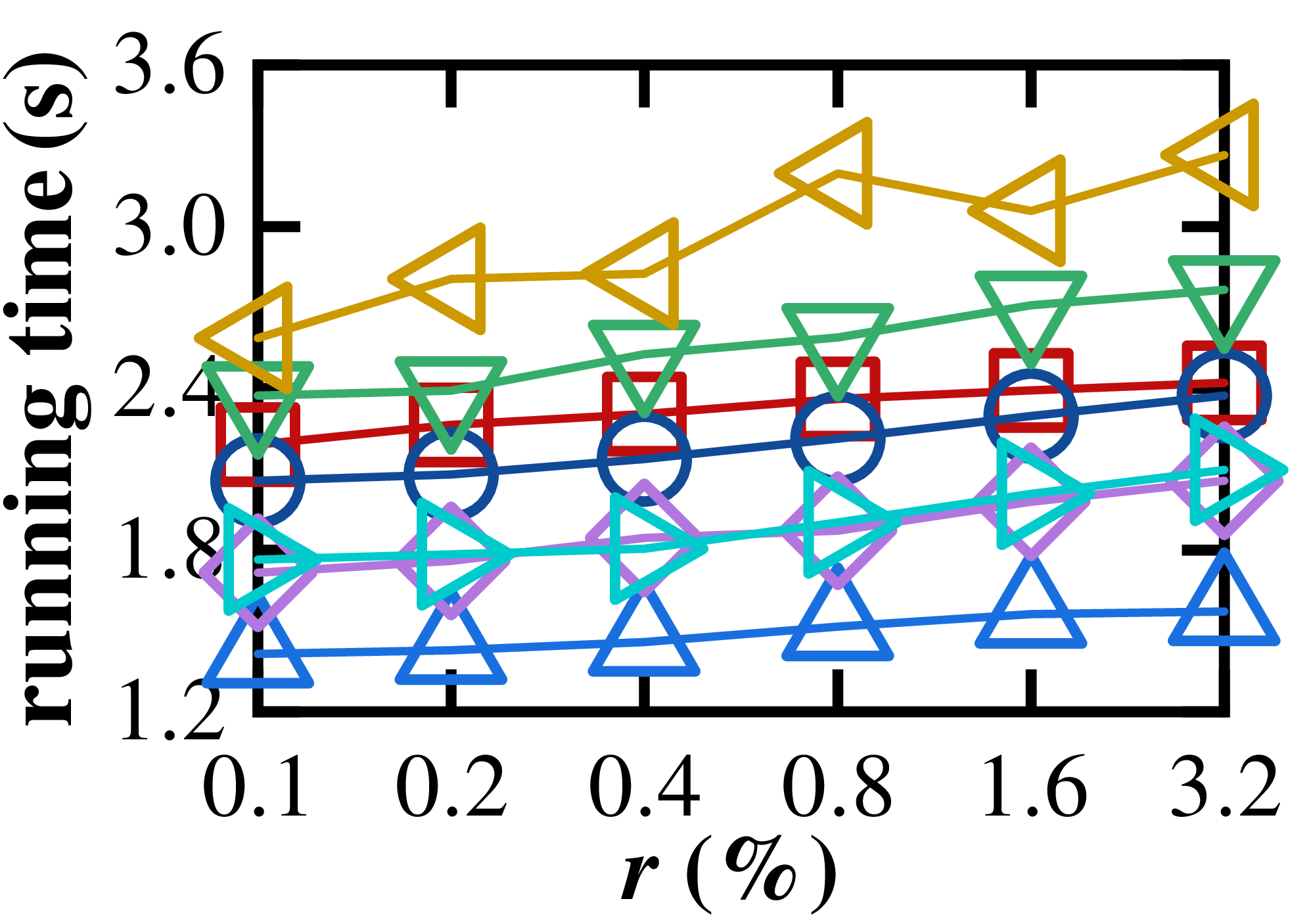}
}
\subfigure[\textit{T-Loc}]{
  \includegraphics[width=0.225\textwidth,height=0.14\textwidth]{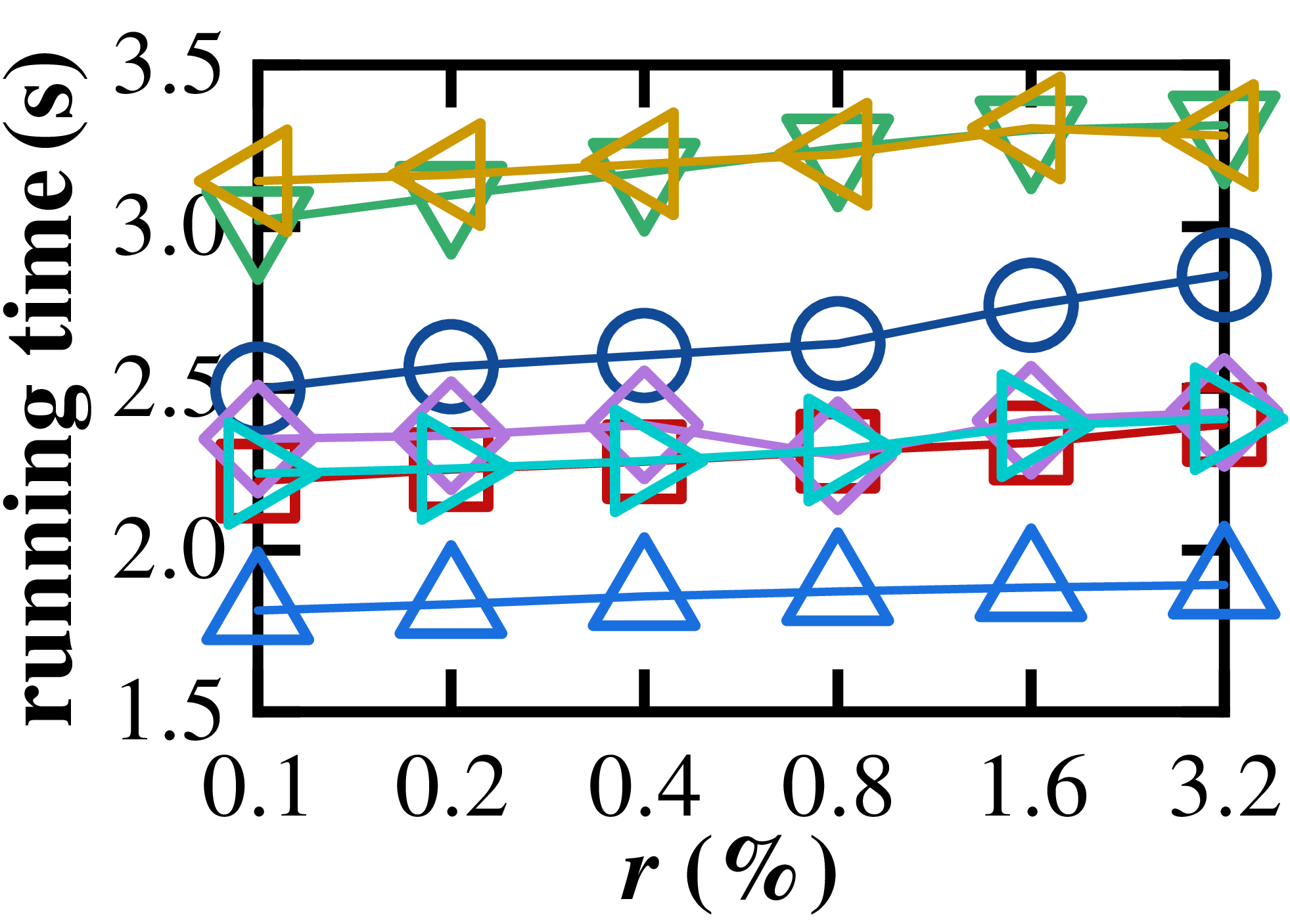}
}\vspace{-0.2cm}

\subfigure[\textit{Vector}]{
  \includegraphics[width=0.225\textwidth,height=0.14\textwidth]{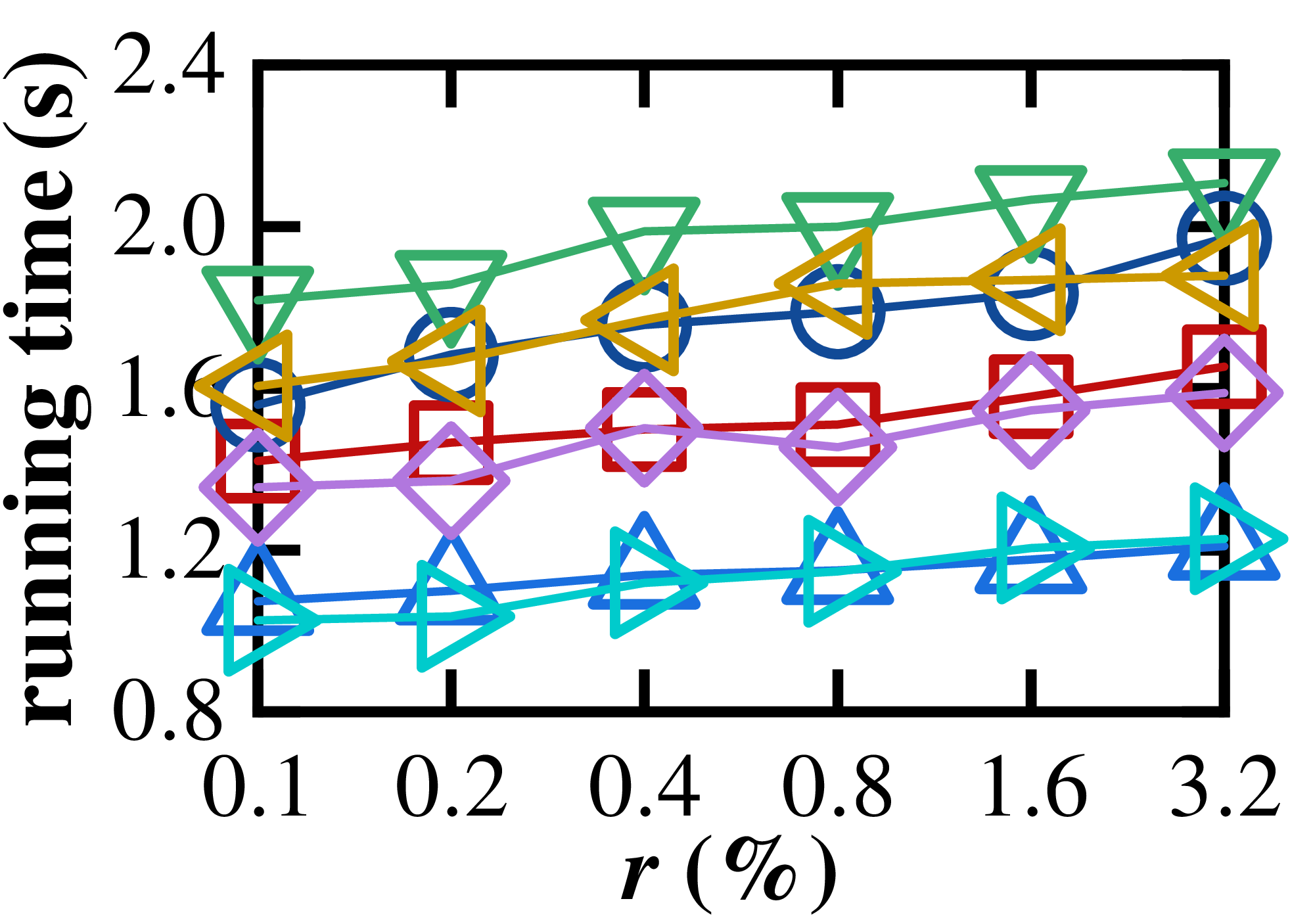}
}
\subfigure[\textit{Color}]{
  \includegraphics[width=0.225\textwidth,height=0.14\textwidth]{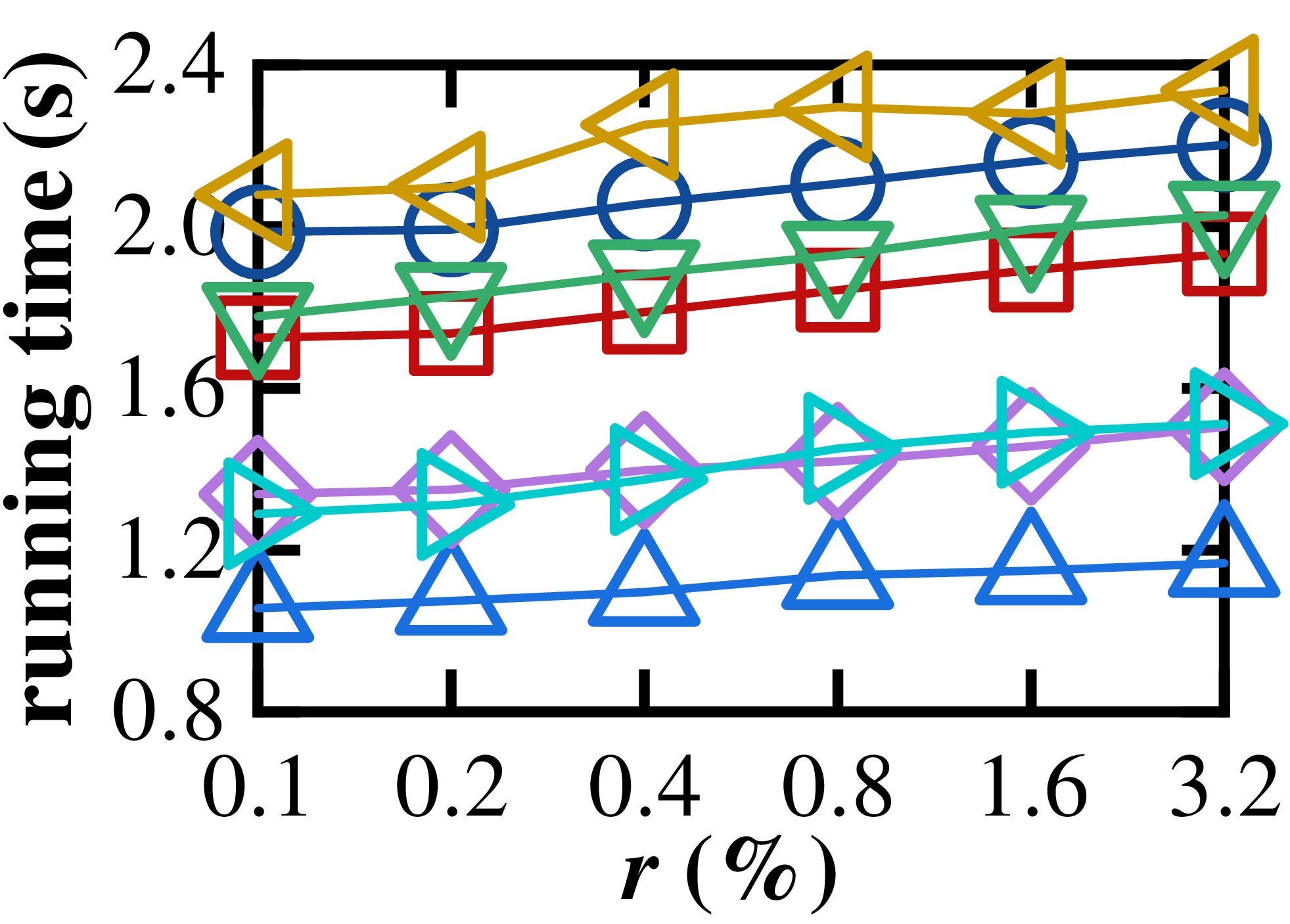}
}
\vspace{-0.5cm}
\caption{\rev{Effect of the search radius $r$}}
\vspace{-0.3cm}
\label{fig:rnn}
\end{center}
\end{figure}

\begin{figure}[t]
\begin{center}
\subfigtopskip=-10pt
\subfigcapskip=-4pt
\includegraphics[height=0.78cm]{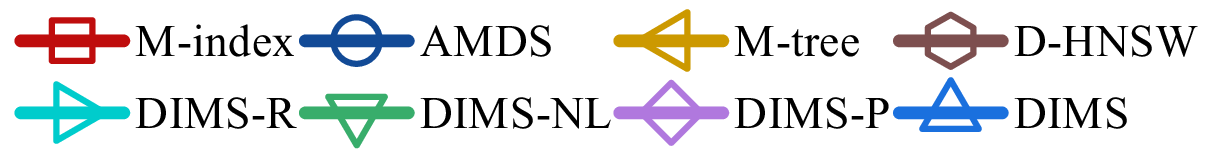}\vspace{-0.05cm}

\subfigure[\textit{Words}]{
  \includegraphics[width=0.225\textwidth,height=0.14\textwidth]{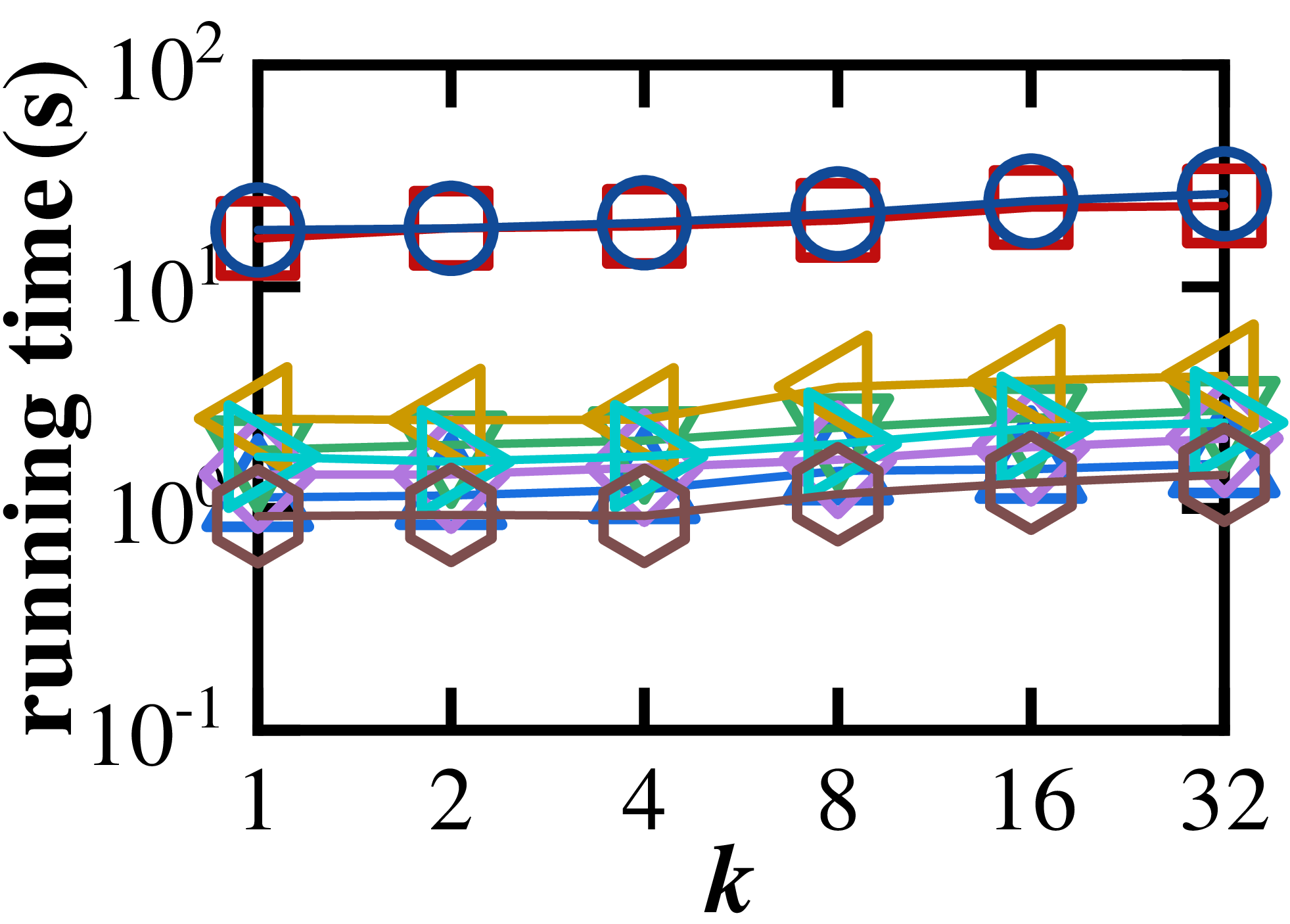}
}
\subfigure[\textit{T-Loc}]{
  \includegraphics[width=0.225\textwidth,height=0.14\textwidth]{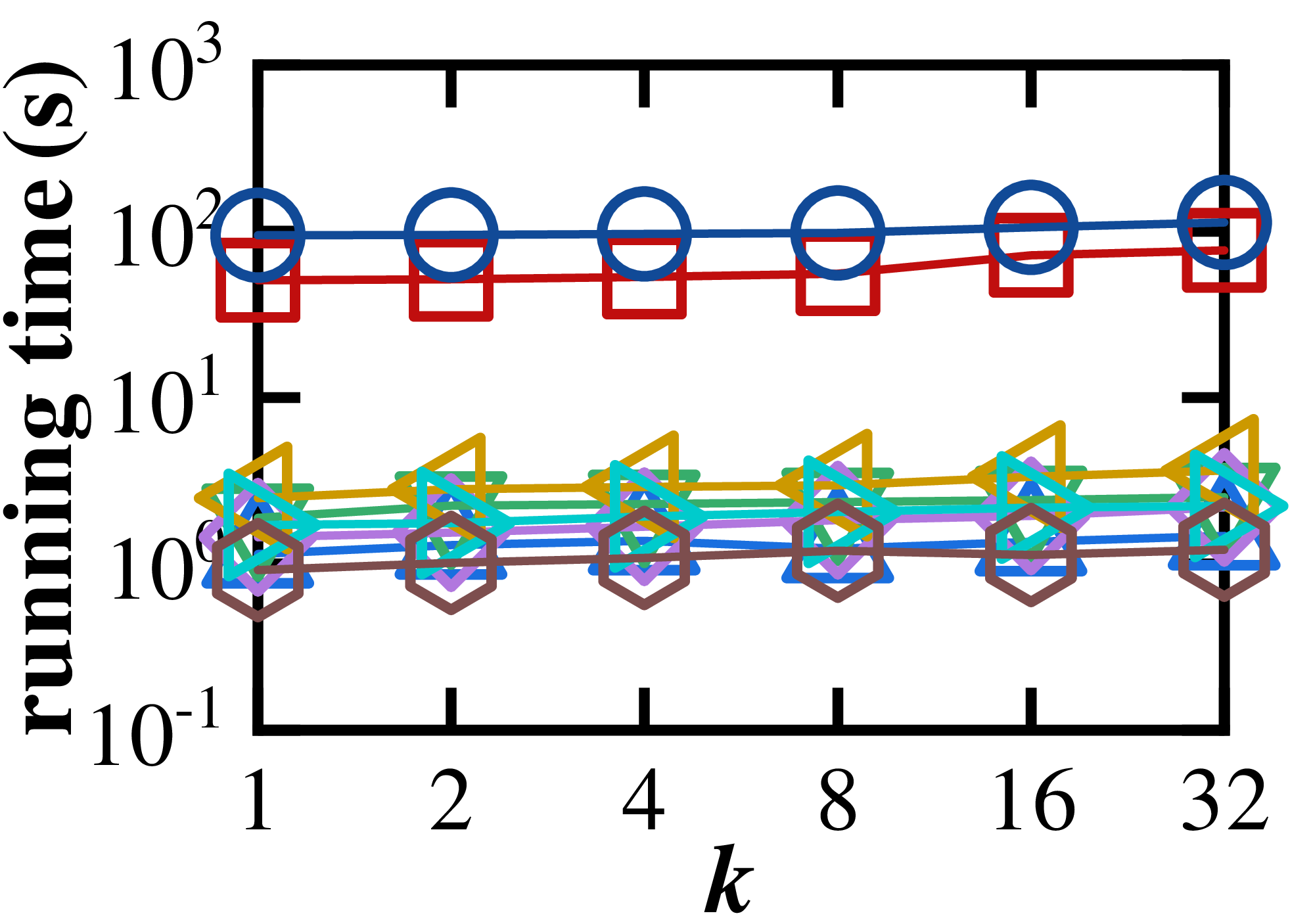}
}\vspace{-0.2cm}

\subfigure[\textit{Vector}]{
  \includegraphics[width=0.225\textwidth,height=0.14\textwidth]{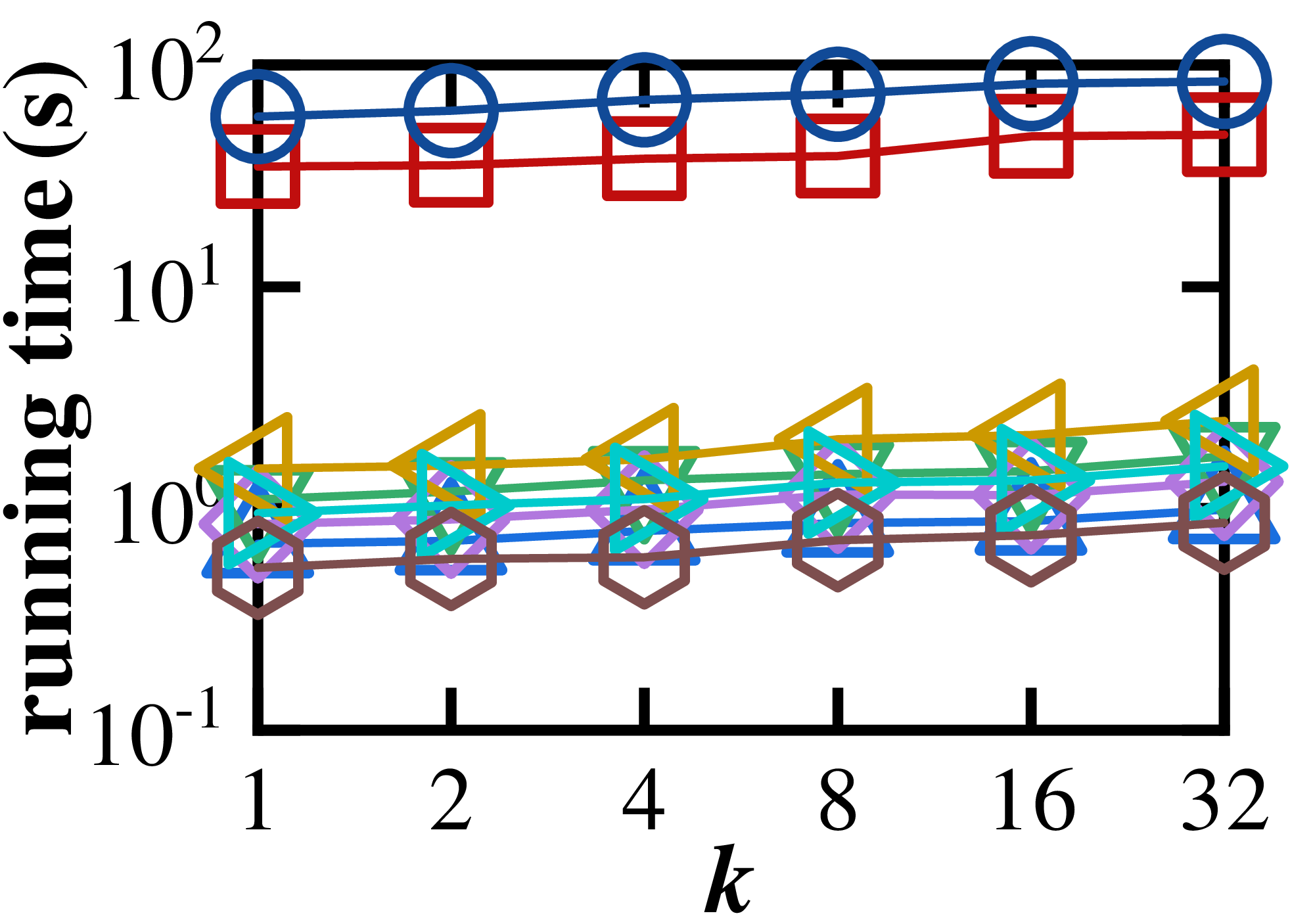}
}
\subfigure[\textit{Color}]{
  \includegraphics[width=0.225\textwidth,height=0.14\textwidth]{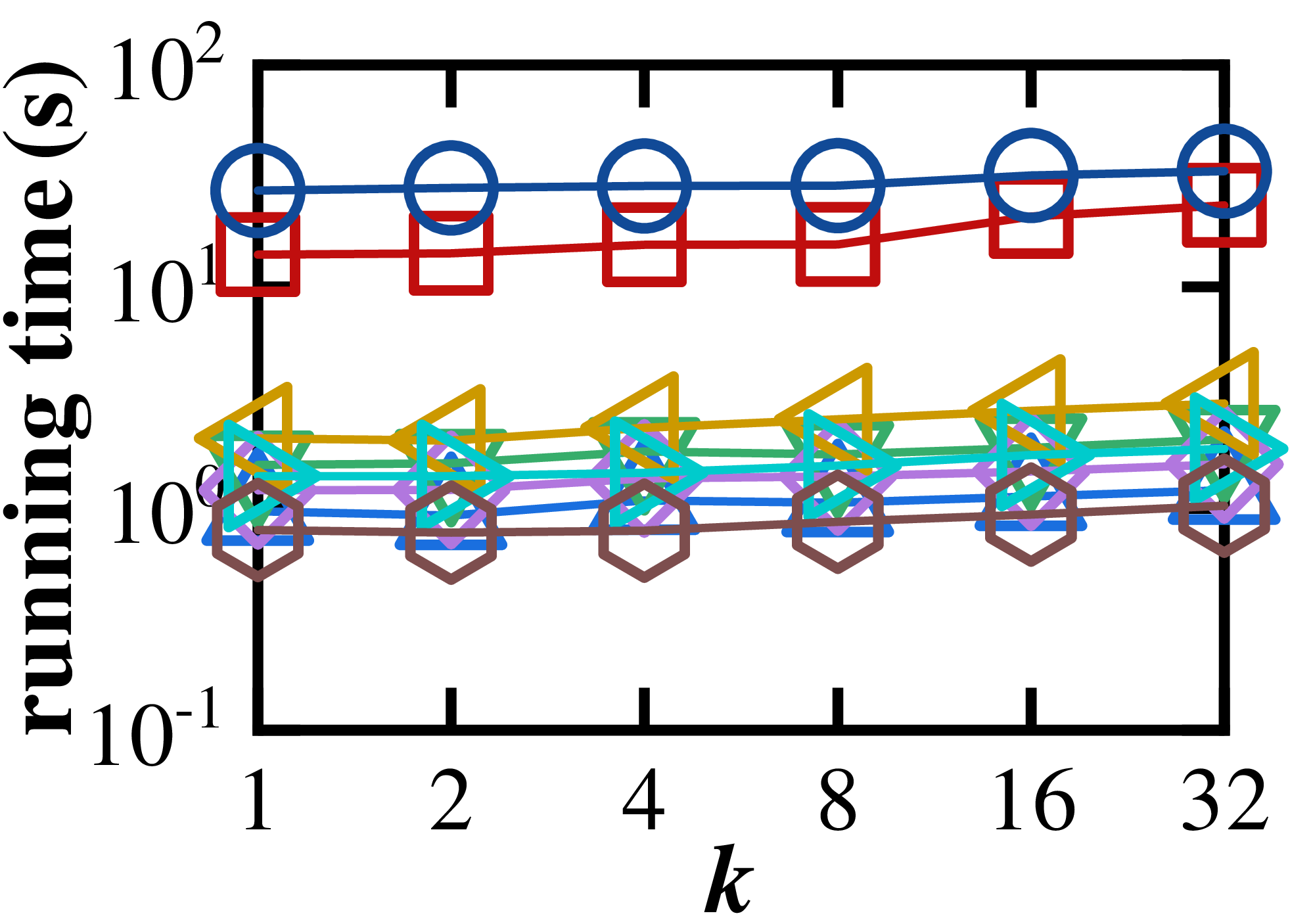}
}
\vspace{-0.4cm}
\caption{\rev{Effect of the integer $k$}}
\vspace{-0.3cm}
\label{fig:knnn}
\end{center}
\end{figure}

\begin{figure}[t]
\begin{center}
\subfigtopskip=-7pt
\subfigcapskip=-3pt
\includegraphics[height=0.44cm]{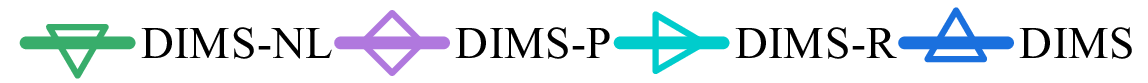}\vspace{0.015cm}

\subfigure[\textit{T-Loc}]{  
  \includegraphics[width=0.225\textwidth,height=0.14\textwidth]{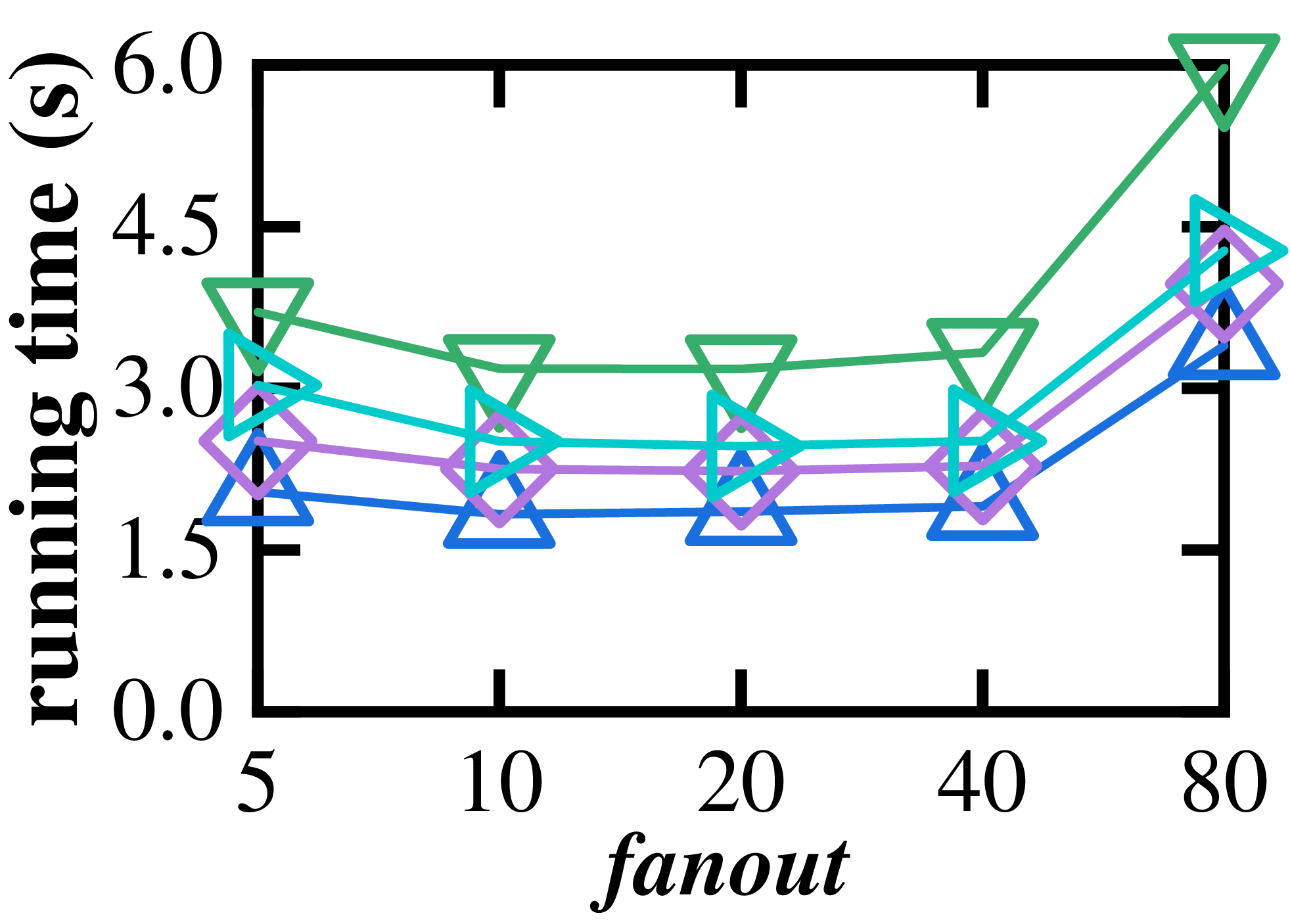}  
}
\subfigure[\textit{Color}]{  
  \includegraphics[width=0.225\textwidth,height=0.14\textwidth]{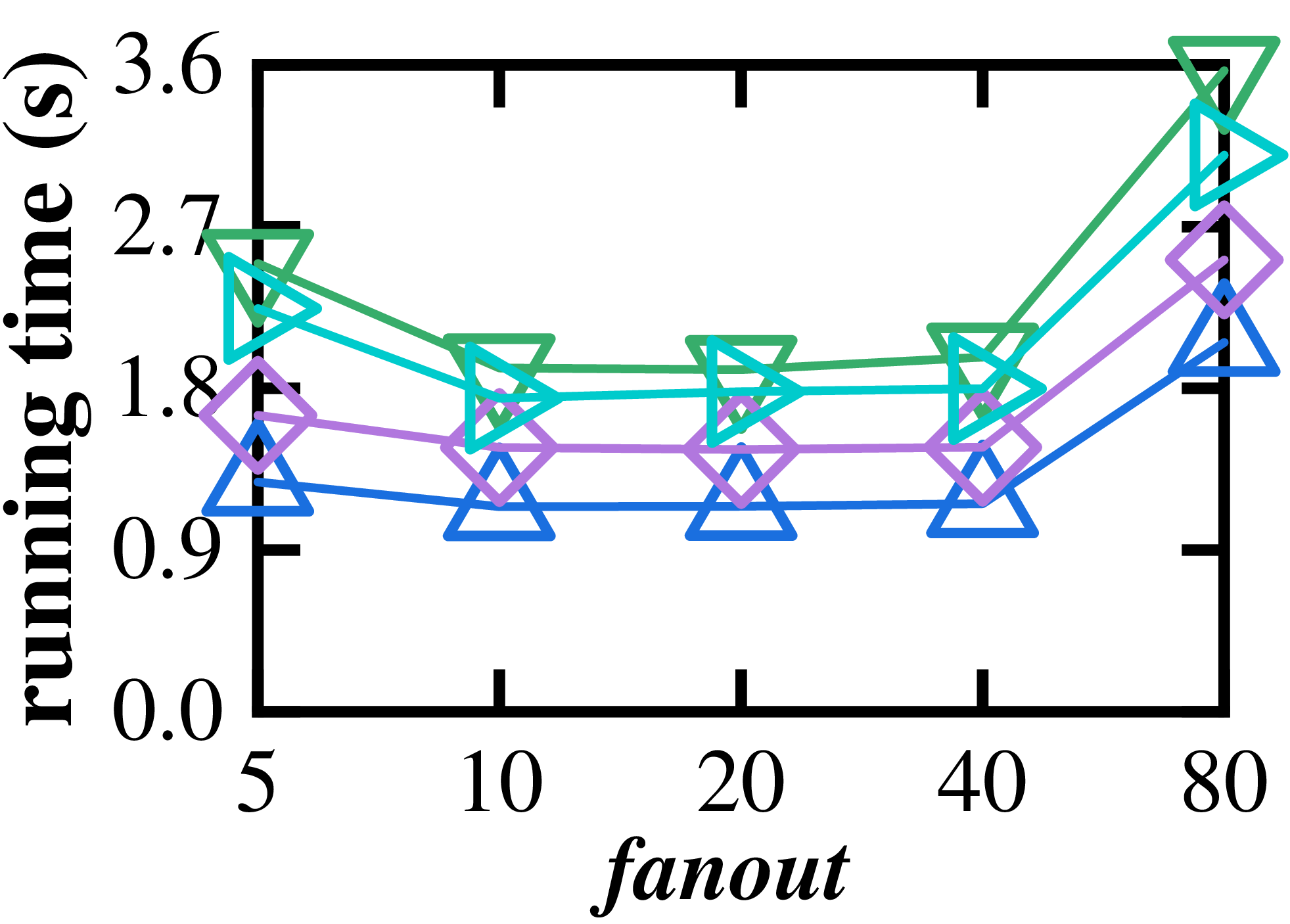}
}
\vspace{-0.4cm}
\caption{\rev{Effect of the M-tree node fanout on MRQ}}
\vspace{-0.3cm}
\label{fig:exp-fanoutM}
\end{center}
\end{figure}

\begin{figure}[t]
\begin{center}
\subfigtopskip=-7pt
\subfigcapskip=-3pt
\includegraphics[height=0.44cm]{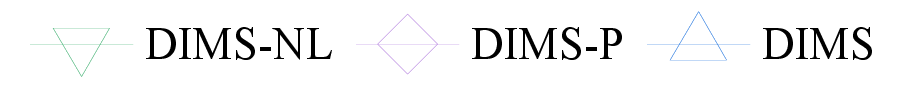}\vspace{0.015cm}

\subfigure[\textit{T-Loc}]{  
  \includegraphics[width=0.225\textwidth,height=0.14\textwidth]{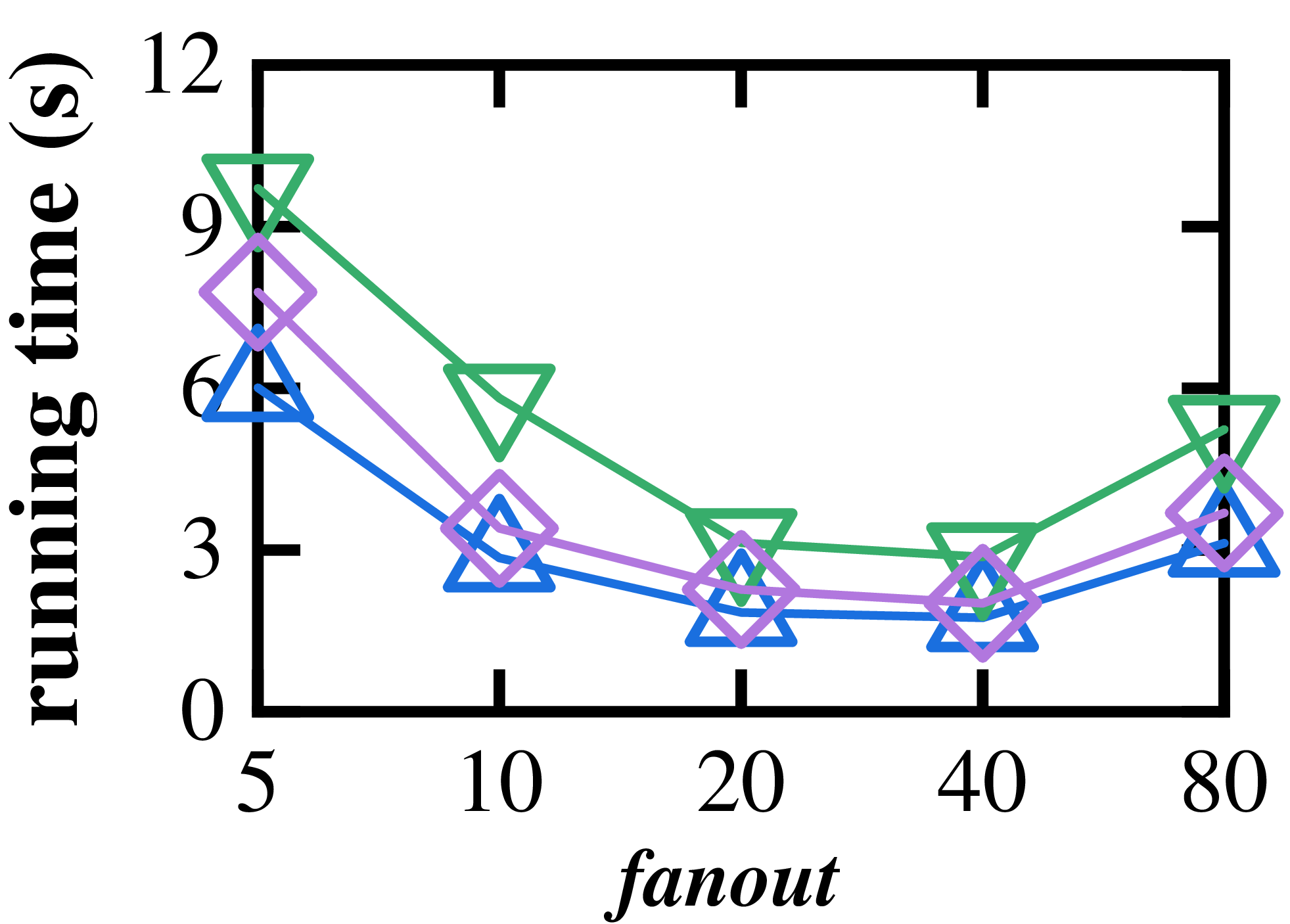}  
}
\subfigure[\textit{Color}]{  
  \includegraphics[width=0.225\textwidth,height=0.14\textwidth]{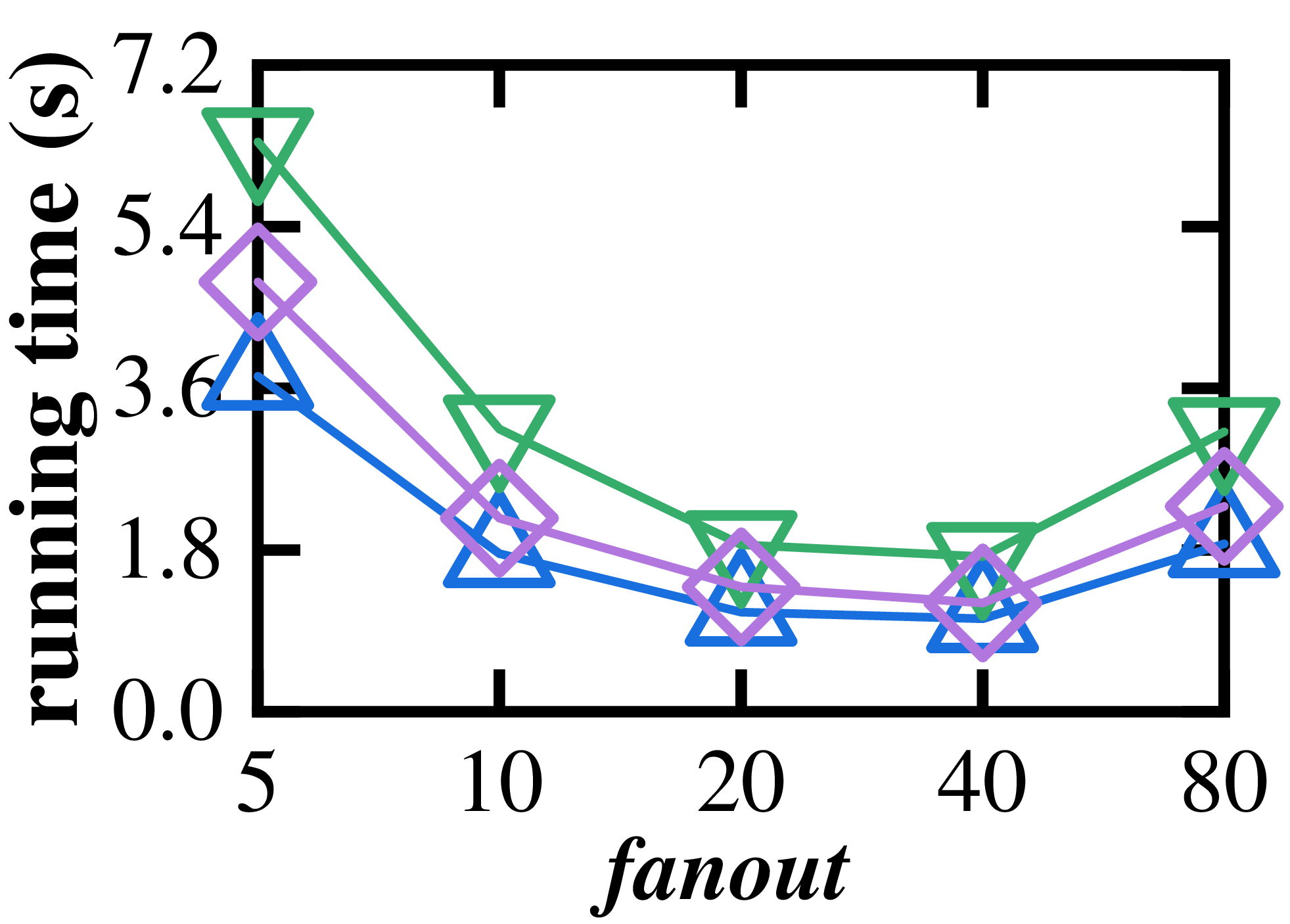}
}
\vspace{-0.4cm}
\caption{\rev{Effect of the B$^+$-tree node fanout on MRQ}}
\vspace{-0.3cm}
\label{fig:exp-fanoutB}
\end{center}
\end{figure}

\begin{figure}[t]
\begin{center}
\subfigtopskip=-10pt
\subfigcapskip=-4pt
\includegraphics[height=0.44cm]{ExpFigs/icon-dims.eps}\vspace{0.015cm}

\subfigure[\textit{T-Loc}]{
  \includegraphics[width=0.225\textwidth,height=0.14\textwidth]{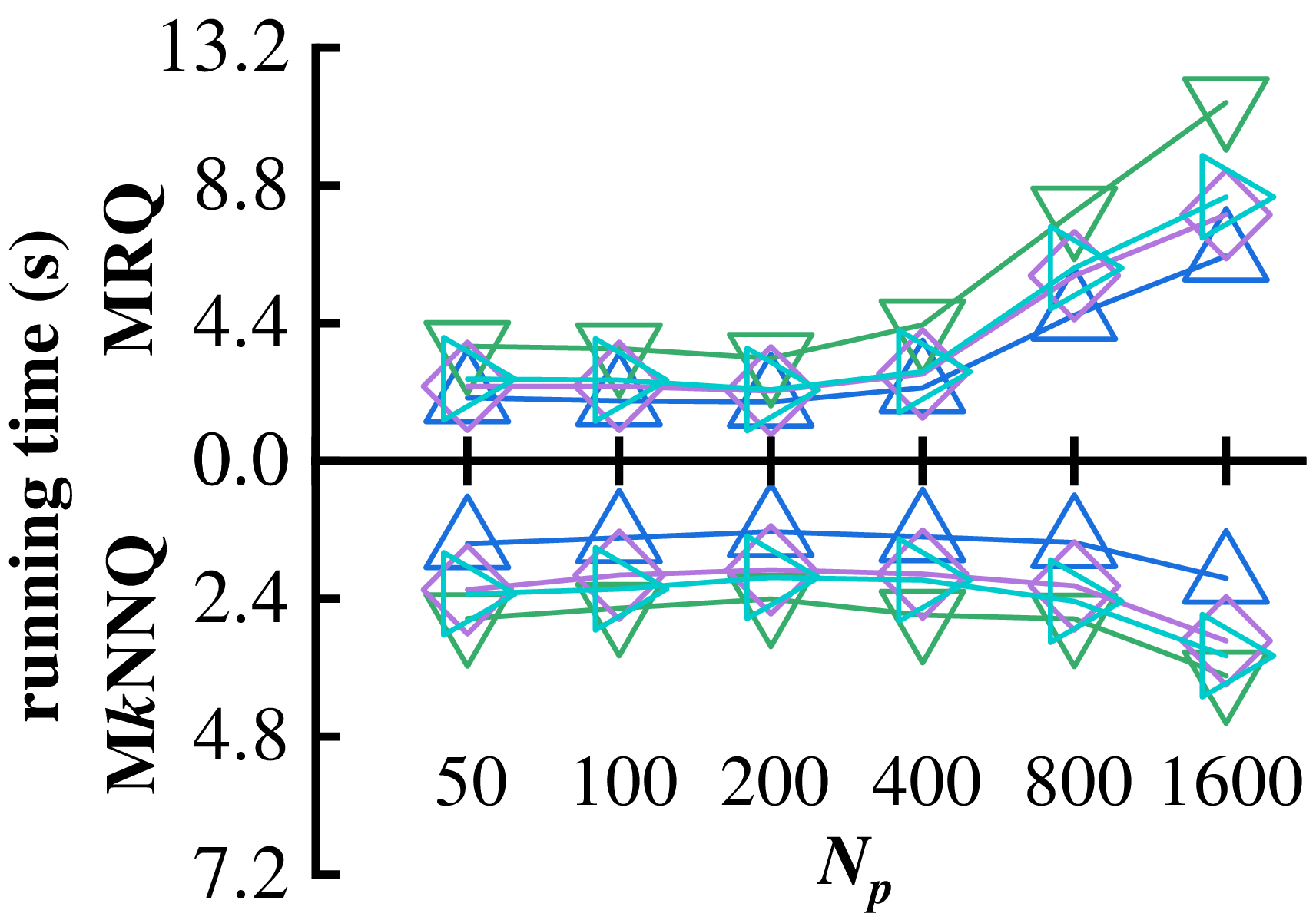} 
}
\subfigure[\textit{Vector}]{
  \includegraphics[width=0.225\textwidth,height=0.14\textwidth]{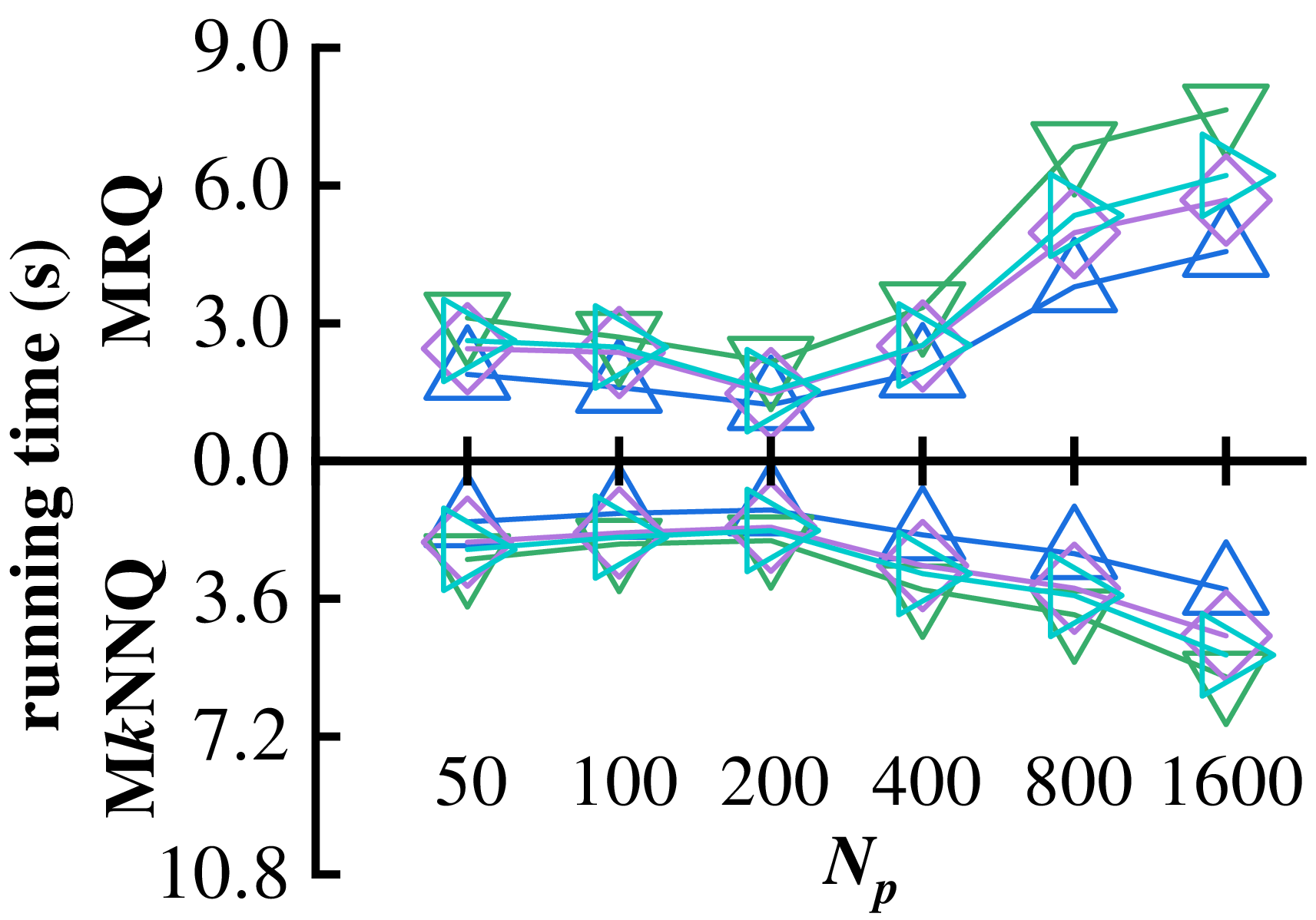} 
}
\vspace{-0.5cm}
\caption{\rev{Effect of the partition number $N_p$ on M\textit{k}NNQ ($N_P^*=200$)}}
\vspace{-0.4cm}
\label{fig:exp-partition}
\end{center}
\end{figure}

\begin{figure}[t]
\begin{center}
\subfigtopskip=-10pt
\subfigcapskip=-4pt
\includegraphics[height=0.48cm]{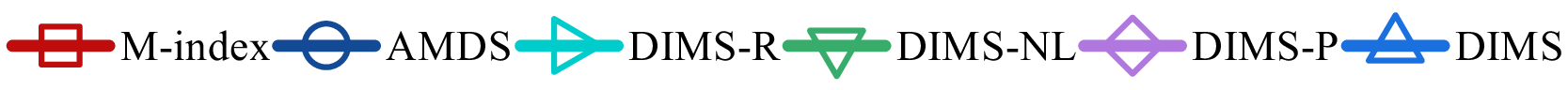}\vspace{-0.08cm}

\subfigure[\textit{Words}]{

  \includegraphics[width=0.225\textwidth,height=0.14\textwidth]{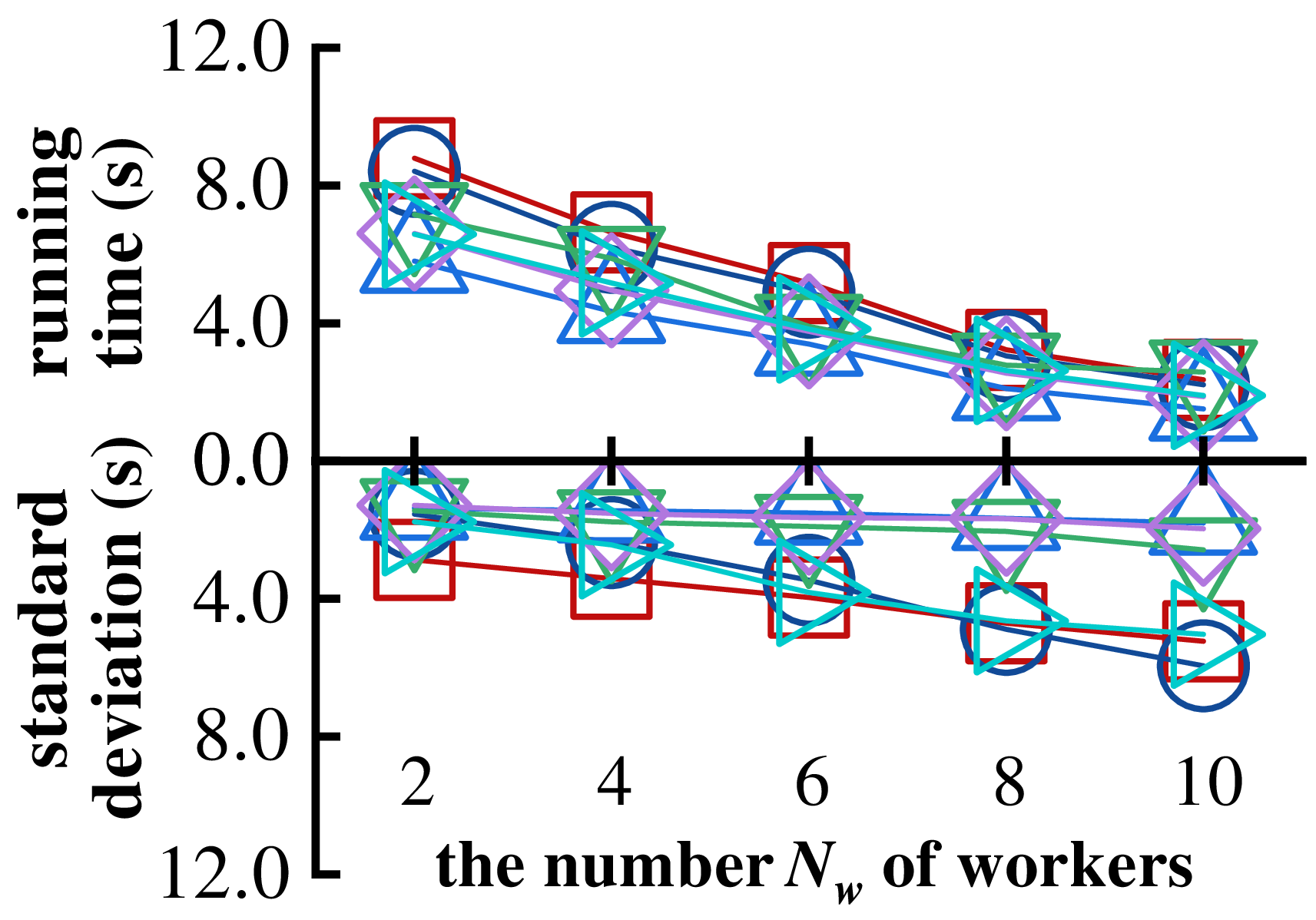}
}\hspace{-0.12cm}
\subfigure[\textit{Color}]{

  \includegraphics[width=0.225\textwidth,height=0.14\textwidth]{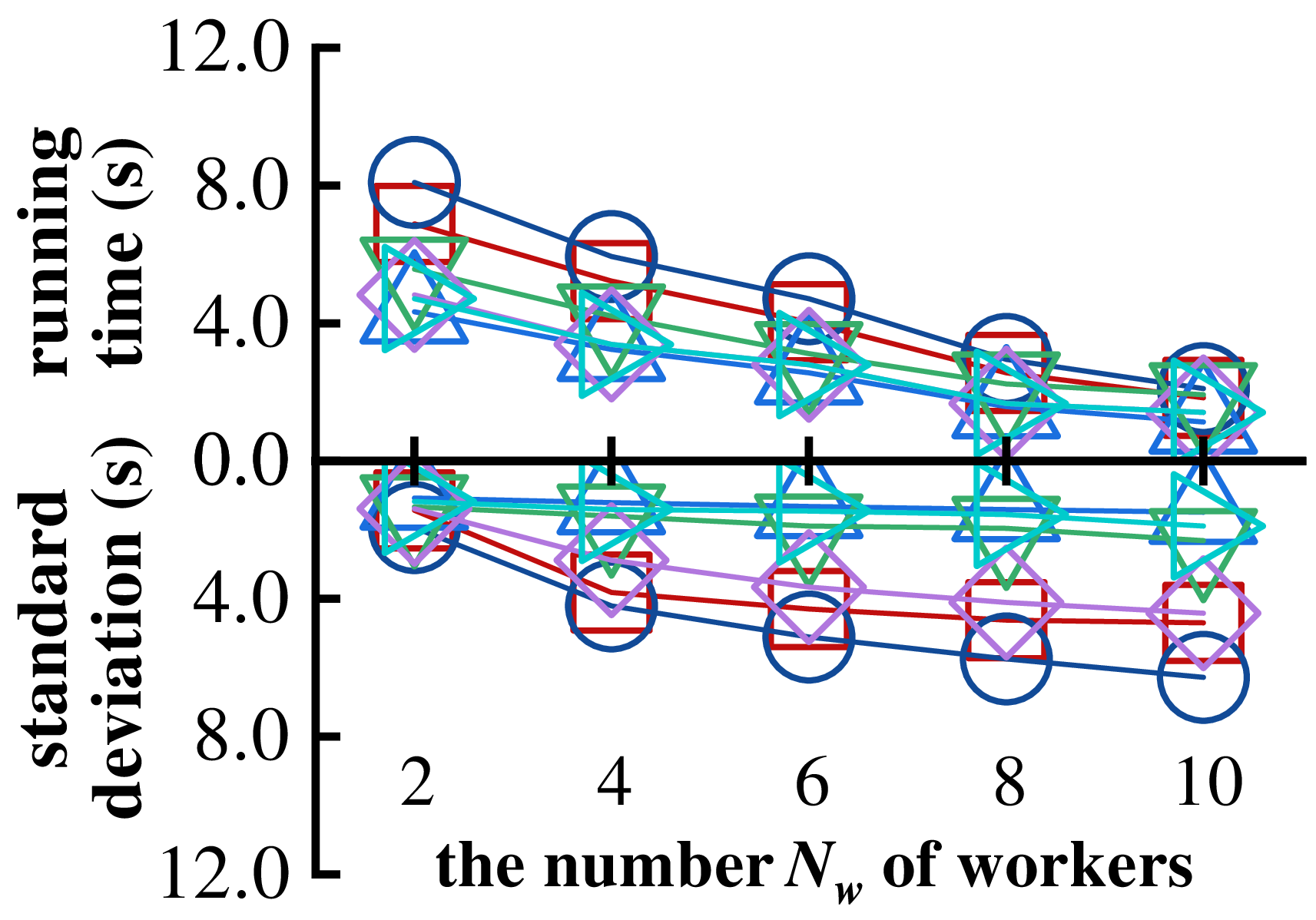}
}
\vspace{-0.3cm}
\caption{\rev{Effect of the number $N_w$ of workers on MRQ }}
\vspace{-0.3cm}
\label{fig:worker}
\end{center}
\end{figure}

\begin{figure}[t]
\begin{center}
\subfigtopskip=-7pt
\subfigcapskip=-4pt
\includegraphics[height=0.78cm]{ExpFigs/icon.eps}\vspace{-0.010cm}
\subfigure[\textit{Words}]{
  \includegraphics[width=0.225\textwidth,height=0.14\textwidth]{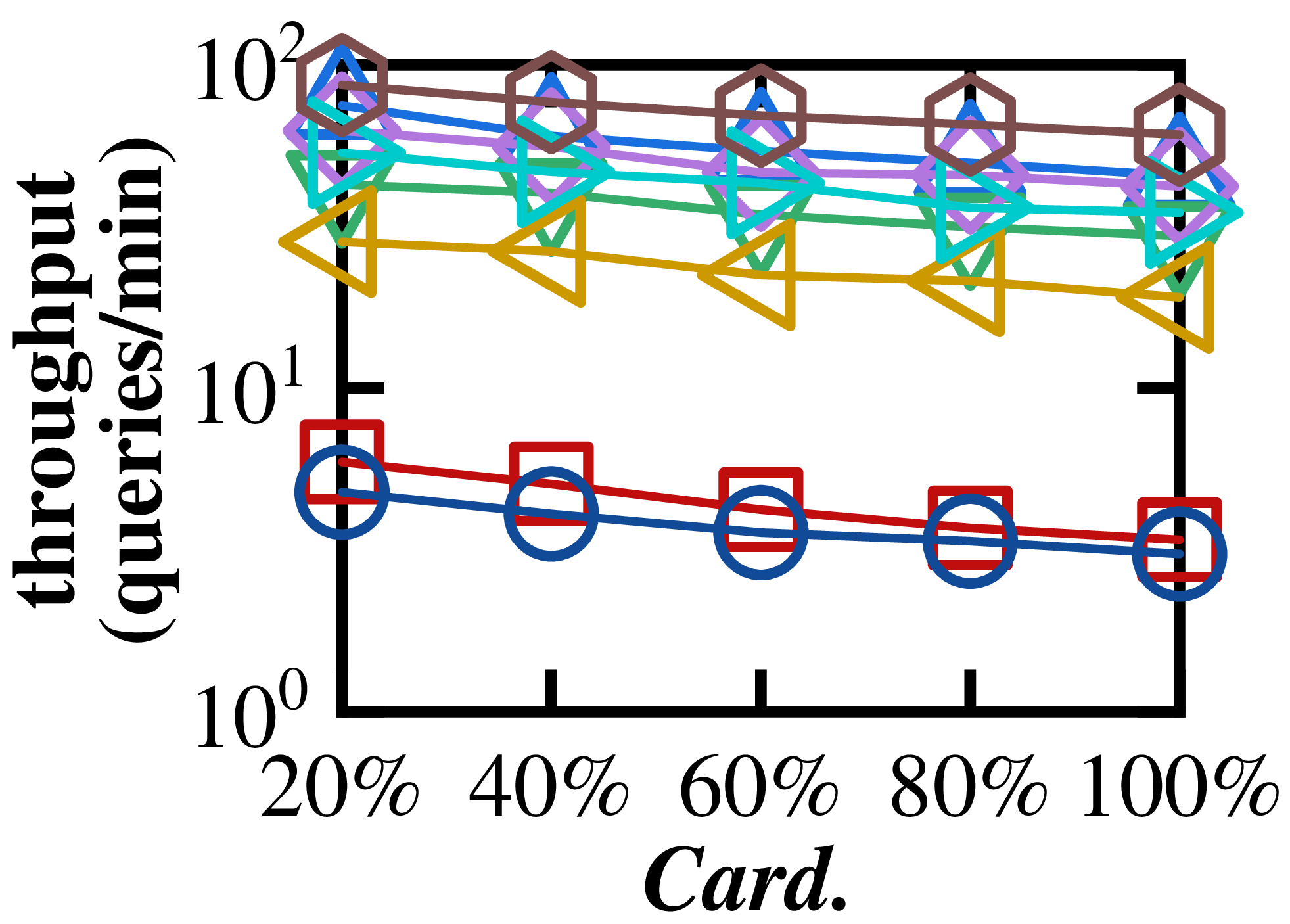}
}
\subfigure[\textit{T-Loc}]{
  \includegraphics[width=0.225\textwidth,height=0.14\textwidth]{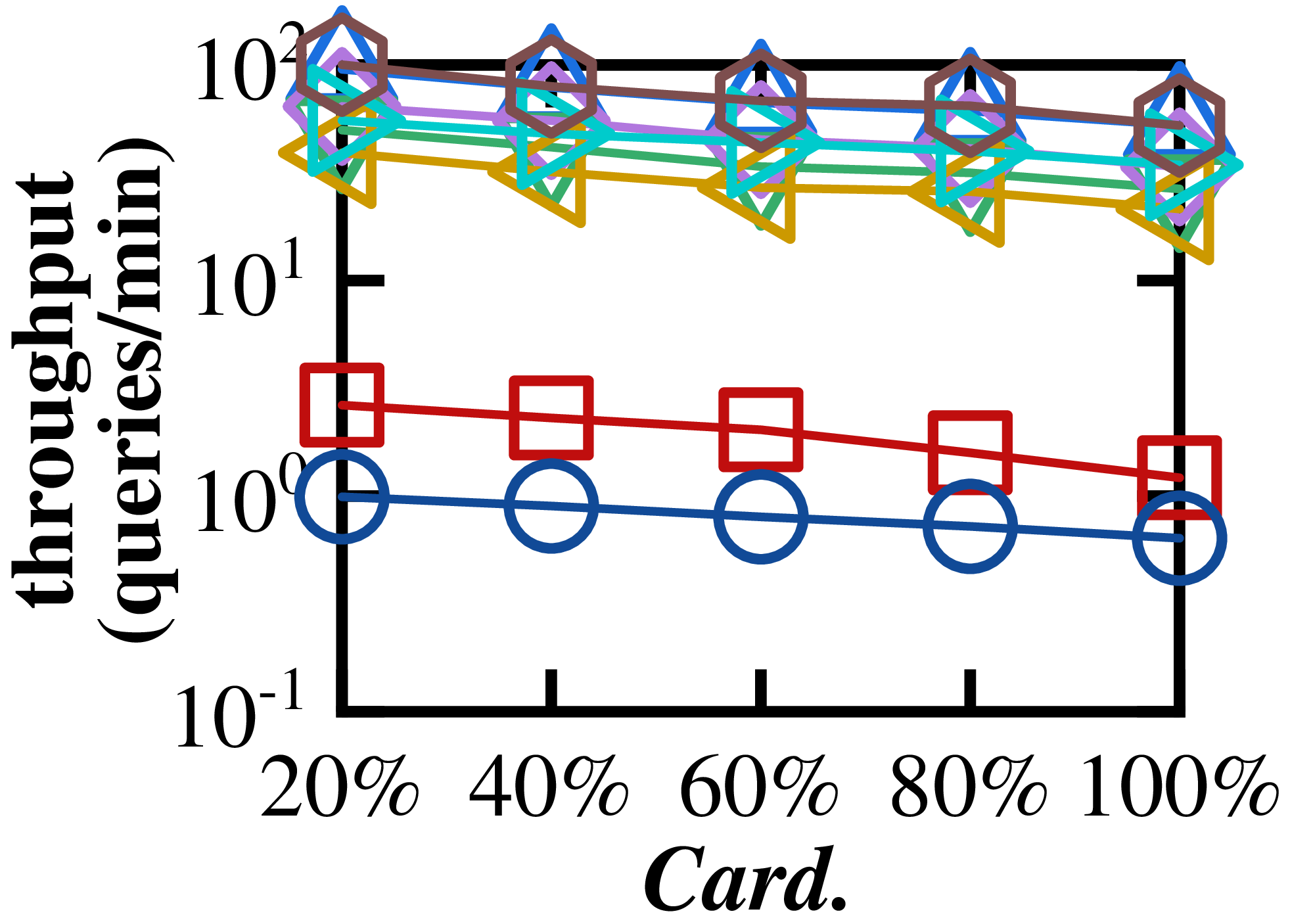}
}
\vspace{-0.1cm}

\subfigure[\textit{Vector}]{
  \includegraphics[width=0.225\textwidth,height=0.14\textwidth]{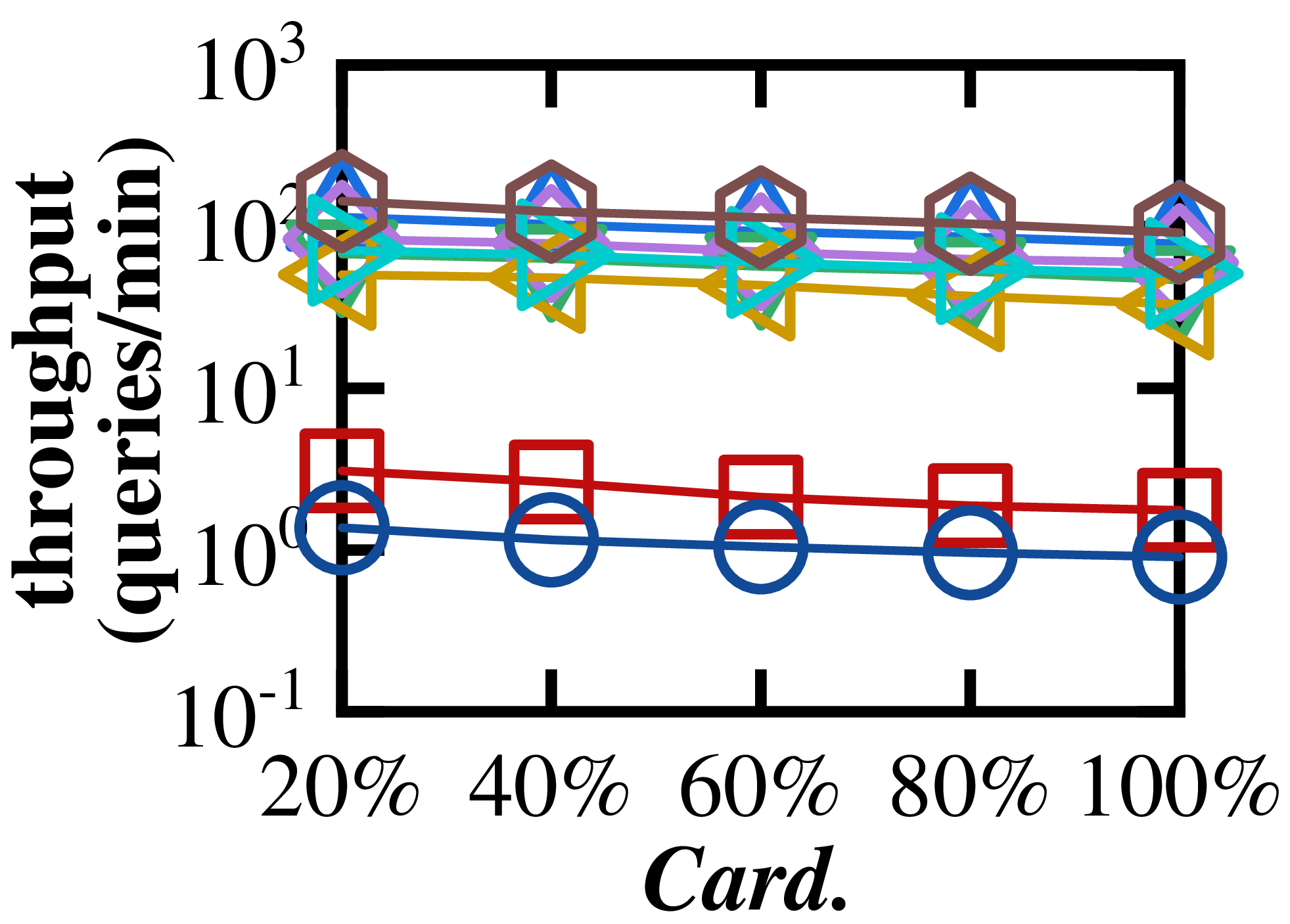}
}
\subfigure[\textit{Color}]{
  \includegraphics[width=0.225\textwidth,height=0.14\textwidth]{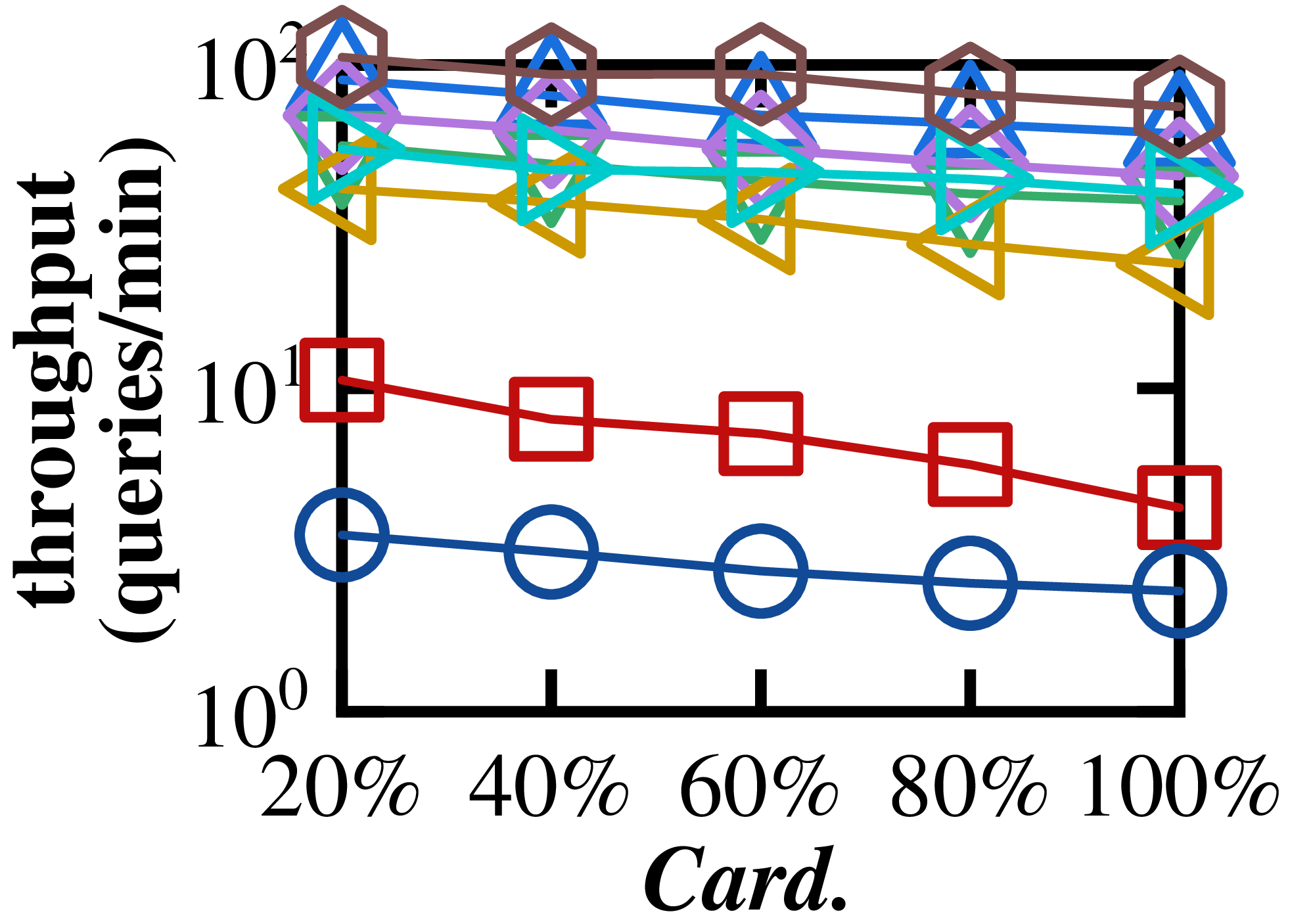}
}

\subfigure[\textit{DEEP}]{
  \includegraphics[width=0.225\textwidth,height=0.14\textwidth]{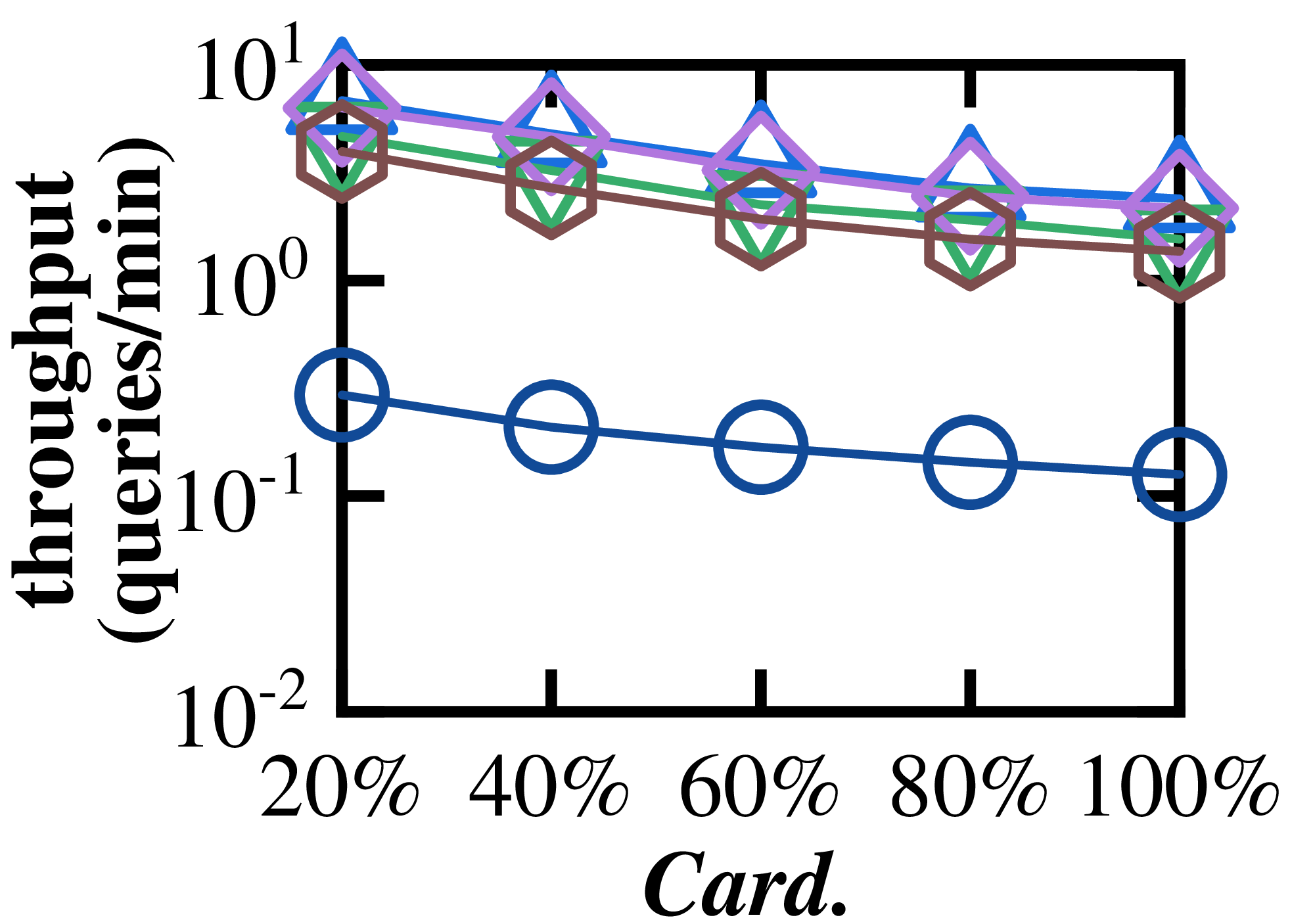}
}
\vspace{-0.3cm}
\caption{\rev{M$k$NNQ performance vs. \textit{Cardinality}}}
\vspace{-0.4cm}
\label{fig:card-knn}
\end{center}
\end{figure}

\vspace{-2mm}
\subsection{Similarity Search Performance}
\vspace{-1mm}
\label{sec:ssp}
We proceed to explore the similarity search performance of DIMS and its competitors by varying three parameters, including the search radius $r$ for MRQ, the desired number $k$ for M$k$NNQ, and the tuning parameter $N_p$.

\noindent
\textbf{Effects of $r$ and $k$.} The impact of $r$ and $k$ can be observed in Figs.~\ref{fig:rnn} and~\ref{fig:knnn} respectively, which illustrate the MRQ and M\textit{k}NNQ performance of different algorithms across four real datasets. 
It's notable that DIMS-NL outperforms M-index and AMDS for M\textit{k}NNQ, while the opposite is observed for MRQ. \rev{Additionally, though D-HNSW achieves the highest efficiency for M\textit{k}NNQ, it fails to support MRQ. In contrast, our proposed DIMS surpasses all general baseline methods, including M-index, AMDs, and M-tree, across all four datasets. This validates the effectiveness and efficiency of our three-stage partitioning technique and the corresponding indexes.} Two main factors contribute to DIMS's superiority.
Firstly, in distributed settings, conducting similarity searches in each worker node and consolidating outcomes at the primary node is crucial for efficient M$k$NNQ. \rev{DIMS-R, DIMS-NL, DIMS-P, and DIMS, by pre-setting query distance constraints via the primary index, locate the nearest partitions of query objects using global and local indexes. These partitions are then searched by multiple local workers, showcasing effective distributed M\textit{k}NNQ. Conversely, existing methods rely solely on local workers for pruning, leading to unnecessary computations and reduced efficiency. Secondly, for distributed metric range queries, only pruning and returning answers within each worker node based on the query radius is required. DIMS-NL faces challenges due to its lack of local indexes with homogeneous partitioning in worker nodes, requiring traversal of all local partitions, resulting in time consumption. 
In contrast, AMDS and M-index directly prune nodes using local indexes, underscoring the importance of establishing locally indexed homogeneous divisions in worker nodes to enhance MRQ efficiency. By leveraging the global index to filter objects and achieve a balanced workload {through heterogeneous partitioning}, DIMS maximizes the utilization of computation resources. {Meanwhile, although the data groups managed by the intermediate index are small compared to the whole dataset, the intermediate B$^+$-tree index demonstrates significant search performance improvement, enabling DIMS to consistently outperform DIMS-R.} As a result, DIMS achieves up to 2x faster performance for MRQ and 50x faster for M\textit{k}NNQ {compared to existing general methods, including M-index, AMDS, and M-tree}.}


\noindent
{\textbf{Effect of the node fanout.} } {In DIMS, both the number of entries in each M-tree node and the maximum fanout in the B$^+$-tree are adjustable parameters. Thus, we conduct experiments to evaluate the impact of node fanout. The results presented in {Figs.~\ref{fig:exp-fanoutM} and~\ref{fig:exp-fanoutB}} illustrate that the search performance of DIMS is affected by the fanout setting of both M-tree and B$^+$-tree, while the optimal value varies depending on dataset characteristics. Therefore, for fair comparison with existing methods in this paper, we set the default fanout to 20, aligning with AMDS~\cite{dase/YangDZCZG19}.}

\noindent
\textbf{Effect of {partition number} $N_p$.} Fig.~\ref{fig:exp-partition} illustrates the MRQ and M$k$NNQ performance across various values of $N_p$, {which validates the effectiveness of our proposed cost-based optimization model.} Specifically, we first compute the optimal number of partitions, denoted as $N_p^*$, for each dataset according to Equation~\ref{equaopt}. Then, we evaluate the influence of different $N_p$ values on the search performance of DIMS by adjusting the value of ${N_p}$ from $0.25{N_p^*}$ to $8{N_p^*}$. Our observations reveal a consistent pattern, aligning with the theoretical analysis presented in Section~\ref{sec:cost}. Initially, as $N_p$ increases, the search cost decreases due to enhanced global pruning power resulting from improved object distribution across the primary node and worker nodes. However, beyond the optimal value, further increase of $N_p$ lead to the division of local workers into more \rev{groups}, leading to higher communication costs between the primary node and worker nodes during object transformation. Consequently, the search cost starts to rise. These findings confirm the importance of maintaining the number of partitions within the optimal range (i.e., $N_p^*$) to minimize search costs.

\vspace{-2mm}
\subsection{Scalability Analysis}
\vspace{-1mm}
Finally, we investigate the scalability of DIMS by varying the number of workers and dataset cardinality.

\vspace{0.08cm}
\noindent
\textbf{{Effect of the number $N_w$ of workers.}} 
{To demonstrate the effectiveness of our proposed three-stage partition strategy, we compare the MRQ performance and workload variance of DIMS with its competitors while varying the number $N_w$ of workers.} Fig.~\ref{fig:worker} illustrates the running time {(shown in the upper part of each subfigure)} and the standard deviation for worker workloads {(shown in the lower part of each subfigure)}. Our observations are as follows: 
(i) As more workers become available, the standard deviation of workloads tends to increase. This occurs because an higher number of workers increases the likelihood of imbalanced partitioning of objects. For instance, if there are six objects to be verified and only two workers are available, each worker can be assigned three objects. However, with four workers available, it becomes impractical to evenly distribute objects among them.
(ii) 
\rev{As the number of workers $N_w$ increases, the running time of all methods decreases due to the availability of additional computation resources, narrowing the difference in query times between DIMS and existing methods. 
However, on the \textit{Words} dataset, DIMS is only 1.3$\times$ faster than the M-index and AMDS when there are only 2 worker nodes, while it achieves a 2$\times$ speedup when there are 10 worker nodes. This showcases that the performance gap widens with an increasing number of available workers.}
(iii) DIMS consistently exhibits the lowest standard deviation compared to other methods, demonstrating the efficiency of our three-stage partitioning method in achieving a balanced workload and optimizing the utilization of computation resources.

\vspace{0.08cm}
\noindent
\textbf{{Effect of cardinality.}} We adjust the cardinality of all datasets from 20\% to 100\% and present the MRQ and M$k$NNQ results in Fig.~\ref{fig:card-knn}, respectively. It's important to note that throughput is not simply the inverse of running time. Throughput represents the average number of different queries that the system completes per minute, whereas running time refers to the average query cost.  As observed, the throughput decreases linearly with the dataset size. This is because the search space expands as the cardinality grows, resulting in a higher computational cost for pruning and verifying objects. \rev{Notably, DIMS achieves the highest or comparable search efficiency on all datasets and scales effectively with increasing data sizes. Based on these results, we can conclude that the proposed distributed metric index DIMS scales effectively with increasing data sizes.}

\vspace{0.2cm}
\noindent
\textbf{Remark.} \rev{Throughout the entire experiment, the proposed DIMS consistently outperforms general similarity search methods, including single machine methods and distributed metric indexes, and stands out as the optimal solution for M\textit{k}NNQ on \textit{DEEP} and MRQ on all datasets. 
These results suggest that DIMS holds promise for efficiently managing large-scale dynamic datasets with flexible distance metrics in existing distributed applications, such as Spark systems.}


\vspace{-1mm}

\section{conclusions}
\label{sec:conclusion}
In this paper, we propose DIMS, a highly effective distributed index for similarity search in metric spaces. DIMS incorporates a three-stage partition strategy to construct effective global, intermediate, and local indexes, thereby accommodating the diverse characteristics of various data and ensuring a balanced workload distribution. Additionally, we introduce concurrent search methods to facilitate efficient distributed similarity search, while leveraging filtering and validation techniques to minimize unnecessary distance computations. To balance computation and communication costs, we develop a cost-based optimization model. Extensive experiments demonstrate that, compared to state-of-the-art distributed methods, our DIMS offers more efficient similarity search, achieves workload balance, and scales well with data size.
These findings highlight the superior effectiveness and scalability of DIMS, indicating its potential for real-life applications.
Moving forward, we plan to apply learning indexing method and study distributed approximate similarity search to further enhance efficiency.


\bibliographystyle{IEEEtran}
\bibliography{reference}

\newpage

\vspace*{-10ex}
\begin{IEEEbiography}
[\vspace*{-10ex}{\includegraphics[width=0.64in,height=0.85in,clip,keepaspectratio]
{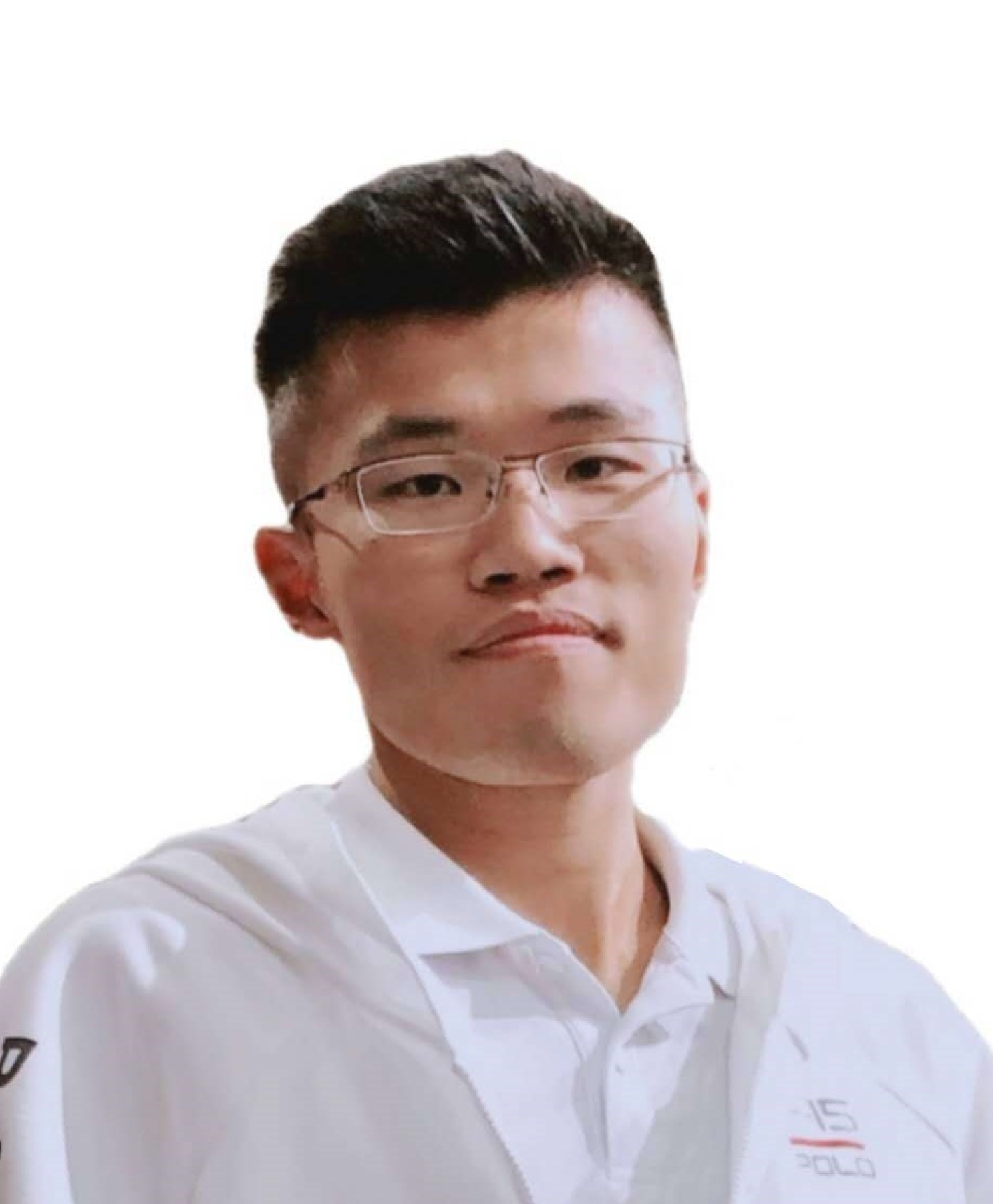}}]{Yifan Zhu}
received the B. S. degree in computer science from Zhejiang University, China, in 2019.
He is currently working toward PhD degree in the College of Computer Science, Zhejiang University, China. His research interests include multi-model data management, vector database management, and hardware acceleration.
\end{IEEEbiography}

\vspace*{-13ex}
\begin{IEEEbiography}[\vspace*{-8ex}{\includegraphics[width=0.64in,height=0.85in,clip,keepaspectratio]
{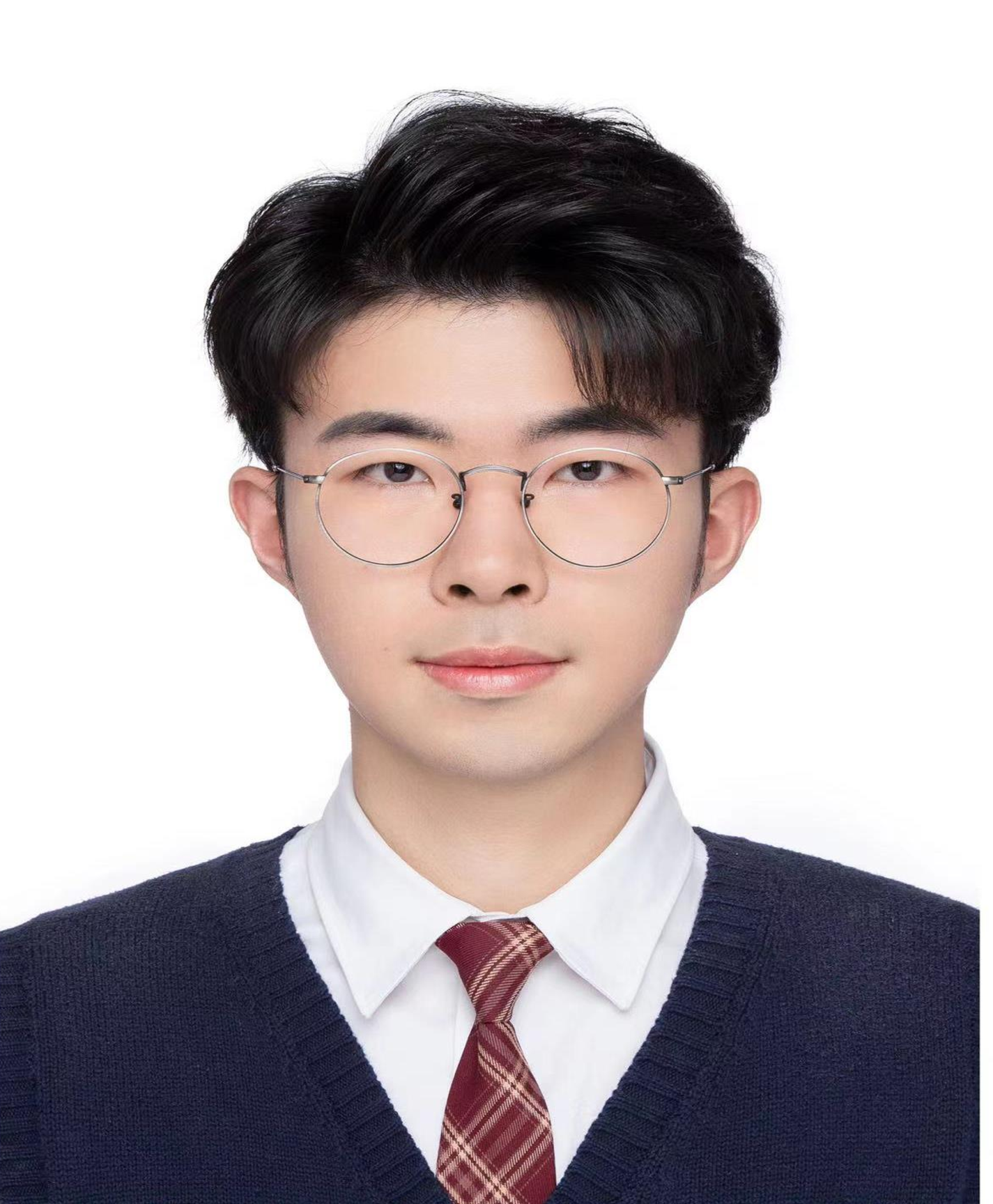}}]{Chengyang Luo}
Chengyang Luo received his B.S. degree in computer science from Nanjing University of Science and Technology, Nanjing, in 2021. He is currently pursuing his M.S. degree in Zhejiang University, Hangzhou. His research interests mainly focus on database usability and graph analysis.
\end{IEEEbiography}

\vspace*{-13ex}
\begin{IEEEbiography}[\vspace*{-8ex}{\includegraphics[width=0.64in,height=0.85in,clip,keepaspectratio]
{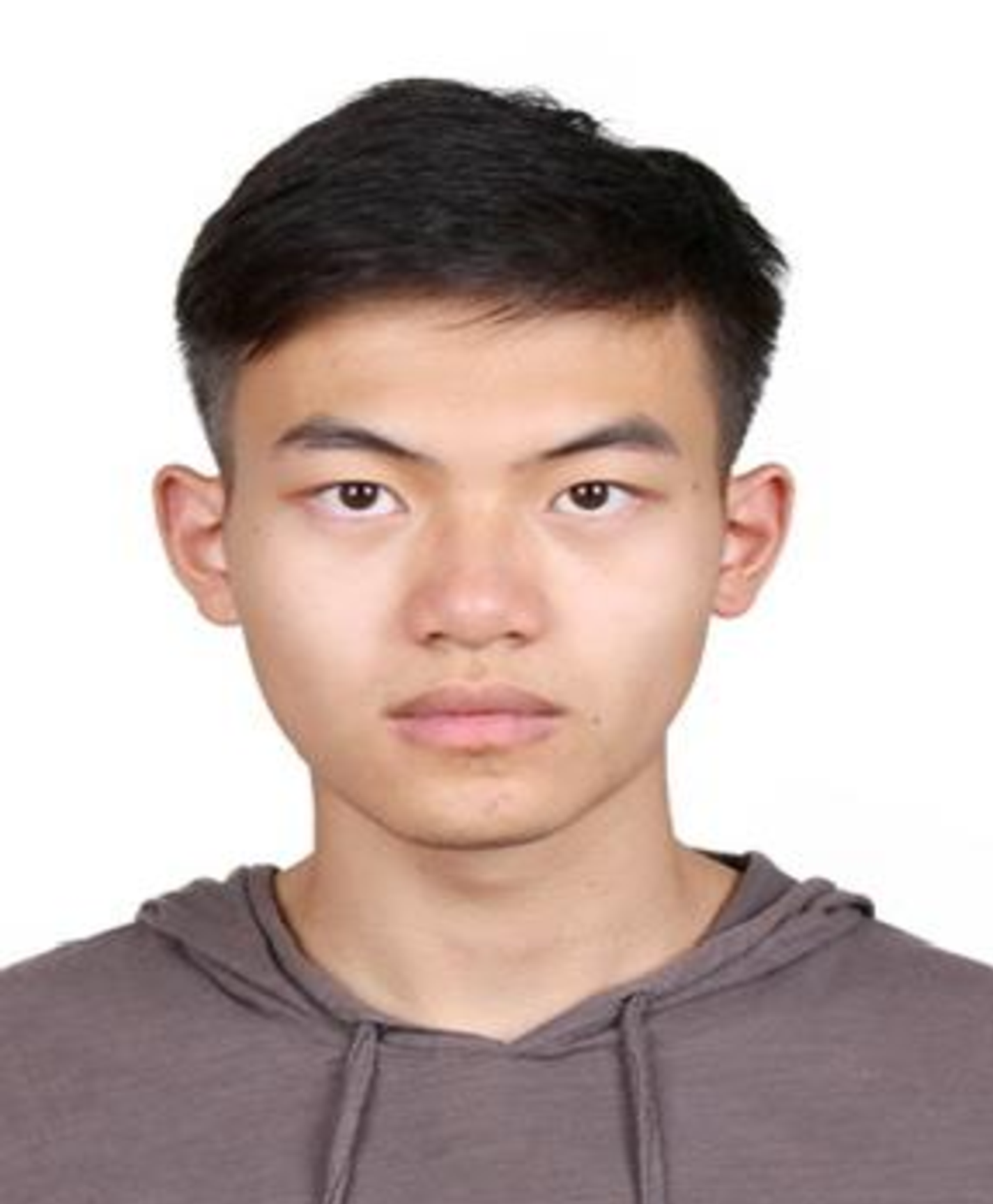}}]{Tang Qian}
Tang Qian received his B.S. degree in computer science from Southwest Jiaotong University, Chengdu, in 2023. He is currently pursuing his M.S. degree in Zhejiang University, Hangzhou. His research interests mainly focus on indexing and querying metric spaces.
\end{IEEEbiography}

\vspace*{-13ex}
\begin{IEEEbiography}[\vspace*{-8ex}{\includegraphics[width=0.64in,height=0.85in,clip,keepaspectratio]
{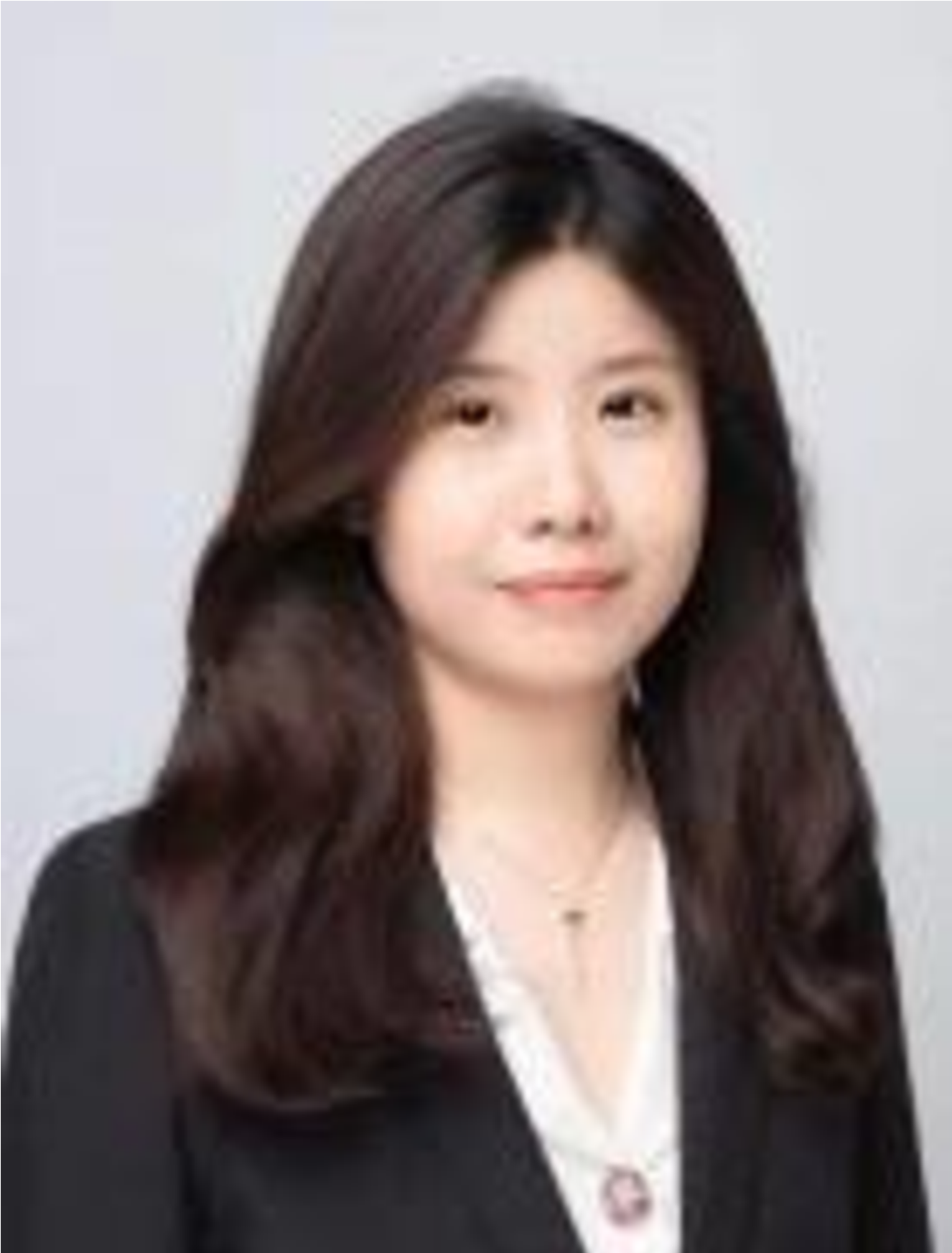}}]{Lu Chen}
received the PhD degree in computer science from Zhejiang University, China, in 2016. She is currently a professor in the College of Computer Science, Zhejiang University, China. Her research interests include indexing and querying metric spaces.
\end{IEEEbiography}

\vspace*{-13ex}
\begin{IEEEbiography}[\vspace*{-8ex}{\includegraphics[width=0.64in,height=0.85in,clip,keepaspectratio]
{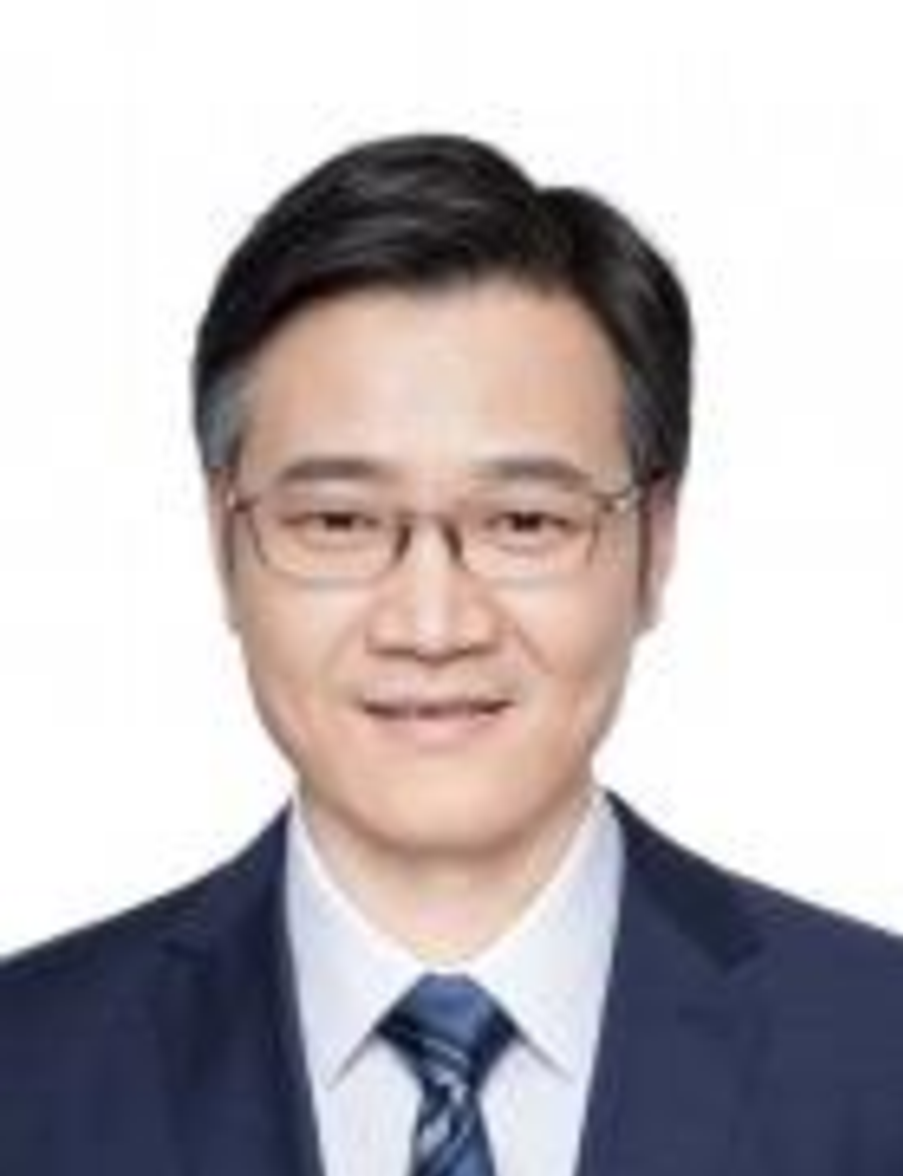}}]{Yunjun Gao} (Senior Member, IEEE)
received the PhD degree in computer science from Zhejiang University, China, in 2008. He is currently a professor in the College of Computer Science, Zhejiang University, China. His research interests include database, Big Data management and analytics, and AI interaction with DB technology. 
\end{IEEEbiography}

\vspace*{-10ex}
\begin{IEEEbiography}[\vspace*{-8ex}{\includegraphics[width=0.64in,height=0.85in,clip,keepaspectratio]{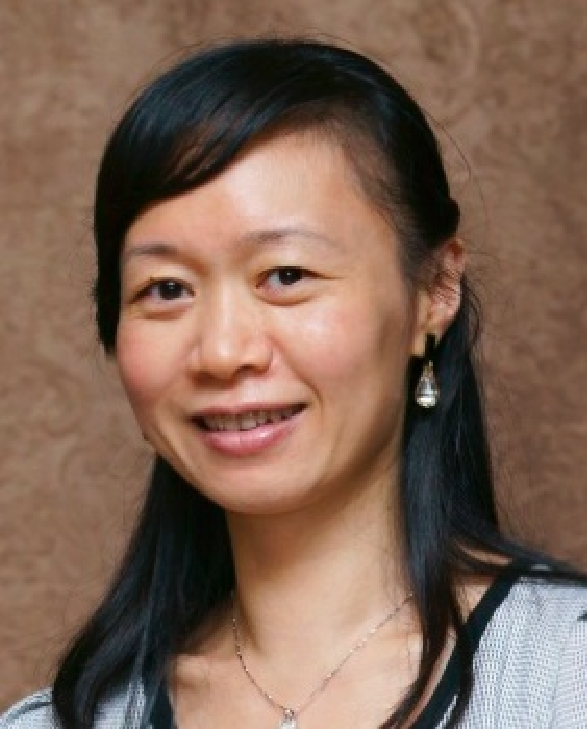}}]{Baihua Zheng}
received her PhD degree in computer science from Hong Kong University of Science \& Technology, China, in 2003. She is currently a Professor in the School of Computing and Information Systems, Singapore Management University, Singapore. Her research interests include mobile/pervasive computing, spatial databases, and big data analytics.
\end{IEEEbiography}

\vfill

\end{document}